\documentclass[longauth]{aa}  
\usepackage{graphicx}
\usepackage[colorlinks=true,citecolor=blue]{hyperref}
\usepackage{multirow}

\usepackage{txfonts}
\newcommand{\logrhk}{$\rm log\,R^{\prime}_\mathrm{HK}$}
\newcommand{\teff}{\ensuremath{T_{\mathrm{eff}}}\xspace}
\newcommand{\kms}{\ensuremath{\rm{km}\,s^{-1}}\xspace}
\newcommand{\ms}{\ensuremath{\rm{m}\,s^{-1}}\xspace}
\newcommand{\logg}{\ensuremath{\log g}\xspace}

\begin{document}


   \title{The GAPS Programme at TNG L}

   \subtitle{TOI-4515\,b: An eccentric warm Jupiter orbiting a 1.2 Gyr-old G-star\thanks{Based on observations made with the Italian Telescopio Nazionale Galileo (TNG) operated by the Fundaci\'on Galileo Galilei (FGG) of the Istituto Nazionale di Astrofisica (INAF) at the Observatorio del Roque de los Muchachos (La Palma, Canary Islands, Spain).}}

   \author{I. Carleo \inst{1,2,3},
    L. Malavolta\inst{4},
    S. Desidera\inst{5},
    D. Nardiello\inst{4,5},
    S. Wang\inst{6},
    D. Turrini\inst{3},
    A. F. Lanza\inst{7},
    M. Baratella\inst{8},
    F. Marzari\inst{4}, 
    S. Benatti\inst{9},
    K. Biazzo\inst{7},
    A. Bieryla\inst{10},
    R. Brahm\inst{11,12,13}
    M. Bonavita\inst{14},
    K. A.\ Collins\inst{10},
    C. Hellier\inst{15},
    D. Locci\inst{9},
    M. J. Hobson\inst{16,12},
    A. Maggio\inst{9},
    G. Mantovan\inst{4},
    S. Messina\inst{7}
    M. Pinamonti\inst{3},
    J. E. Rodriguez\inst{17},
    A. Sozzetti\inst{3},
    K. Stassun\inst{18},
    X. Y. Wang\inst{6},
    C. Ziegler\inst{19},
    M. Damasso\inst{3},
    P. Giacobbe\inst{3},
    F. Murgas\inst{1,2},
    H. Parviainen\inst{1,2},
    G. Andreuzzi\inst{20,21},
    K. Barkaoui\inst{22,23,1},
    P. Berlind\inst{10},
    A. Bignamini\inst{24},
    F. Borsa\inst{25},
    C. Brice\~{n}o\inst{26},
    M. Brogi\inst{27,3},
    L. Cabona\inst{25},
    M. L. Calkins\inst{10},
    R. Capuzzo-Dolcetta\inst{28},
    M. Cecconi\inst{20},
    K.~D.~Colon\inst{29},
    R. Cosentino\inst{20},
    D. Dragomir\inst{30},
    G. A. Esquerdo\inst{10},
    T. Henning\inst{16},
    A. Ghedina\inst{20},
    R.~F.~Goeke\inst{31},
    R. Gratton\inst{5},
    F. Grau Horta\inst{32},
    A. F.\ Gupta\inst{33,34},
    J. M. Jenkins\inst{35},
    A. Jord\'an\inst{11,12,13},
    C. Knapic\inst{24},
    D. W. Latham\inst{10},
    I. Mireles\inst{30},
    N. Law\inst{36},
    V. Lorenzi\inst{20,1},
    M. B. Lund\inst{37},
    J. Maldonado\inst{9},
    A. W. Mann\inst{37},
    E. Molinari\inst{25},
    E. Pall{\'e}\inst{1,2},
    M. Paegert\inst{10},
    M. Pedani\inst{20},
    S. N. Quinn\inst{10},
    G. Scandariato\inst{7},
    S. Seager\inst{31,23,38},
    J. N.\ Winn\inst{39},
    B. Wohler\inst{40,35},  
          \and
    T. Zingales\inst{4}
          }
   \institute{Instituto de Astrof\'isica de Canarias (IAC), Calle V\'ia L\'actea s/n, 38200, La Laguna, Tenerife, Spain 
    \email{ilariacarleo.astro@gmail.com}
        \and Departamento de Astrof\'isica, Universidad de La Laguna (ULL), 38206 La Laguna, Tenerife, Spain 
        \and  INAF - Osservatorio Astrofisico di Torino, Via Osservatorio 20, I-10025 Pino Torinese, Italy 
        \and  Dipartimento di Fisica e Astronomia "Galileo Galilei", Universit{\'a} di Padova, Vicolo dellOsservatorio 3, I-35122, Padova, Italy 
        \and INAF - Osservatorio Astronomico di Padova, Vicolo dell'Osservatorio 5, I-35122 Padova, Italy 
        \and Astronomy Department, Indiana University, Bloomington, IN 47405-7105, USA 
        \and INAF - Osservatorio Astrofisico di Catania, Via S. Sofia 78, I-95123 Catania, Italy 
        \and Leibniz-Institut f{\"u}r Astrophysik Potsdam (AIP), An der Sternwarte 16, D-14482 Potsdam, Germany 
        \and INAF - Osservatorio Astronomico di Palermo, Piazza del Parlamento, 1, I-90134 Palermo, Italy 
        \and Center for Astrophysics \textbar \ Harvard \& Smithsonian, 60 Garden Street, Cambridge, MA 02138, USA 
        \and Facultad de Ingeniera y Ciencias, Universidad Adolfo Ib\'{a}\~{n}ez, Av. Diagonal las Torres 2640, Pe\~{n}alol\'{e}n, Santiago, Chile 
        \and Millennium Institute for Astrophysics, Chile 
        \and Data Observatory Foundation, Chile 
        \and SUPA, Institute for Astronomy, University of Edinburgh, Blackford Hill, Edinburgh EH9 3HJ, UK 
        \and Astrophysics Group, Keele University, Staffs ST5 5BG, U.K.  
        \and Max-Planck-Institut f\"{u}r Astronomie, K\"{o}nigstuhl  17, 69117 Heidelberg, Germany 
        \and Center for Data Intensive and Time Domain Astronomy, Department of Physics and Astronomy, Michigan State University, East Lansing, MI 48824, USA 
        \and Department of Physics \& Astronomy, Vanderbilt University, Nashville, TN, USA 
        \and Department of Physics, Engineering and Astronomy, Stephen F. Austin State University, 1936 North St, Nacogdoches, TX 75962, USA 
        \and Fundaci{\'o}n Galileo Galilei-INAF, Rambla Jos{\'e} Ana Fernandez P{\'e}rez 7, 38712 Bre{\~n}a Baja, TF, Spain 
        \and INAF - Osservatorio Astronomico di Roma, Via Frascati 33, 00078 Monte Porzio Catone, Italy 
        \and Astrobiology Research Unit, Universit{\'e} de Li{\'e}ge, 19C Al{l\'e}e du 6 Ao\^ut, 4000 Li\'ege, Belgium 
        \and Department of Earth, Atmospheric and Planetary Science, Massachusetts Institute of Technology, 77 Massachusetts Avenue, Cambridge, MA 02139, USA 
        \and INAF - Osservatorio Astronomico di Trieste, via Tiepolo 11, 34143 Trieste 
        \and INAF - Osservatorio Astronomico di Brera, Via E. Bianchi 46, 23807 Merate, Italy 
        \and Cerro Tololo Inter-American Observatory/NSF's NOIRLab, Casilla 603, La Serena, Chile 
        \and Dipartimento di Fisica, Universit{\'a} degli Studi di Torino, via Pietro Giuria 1, I-10125, Torino, Italy 
        \and Dipartimento di Fisica, Sapienza, Universit{\'a} di Roma, P.le Aldo Moro, 5, 00185 - Rome, Italy 
        \and NASA Goddard Space Flight Center, Exoplanets and Stellar Astrophysics Laboratory (Code 667), Greenbelt, MD 20771, USA 
        \and Department of Physics and Astronomy, University of New Mexico, 210 Yale Blvd NE, Albuquerque, NM, USA 
        \and Department of Physics and Kavli Institute for Astrophysics and Space Research, Massachusetts Institute of Technology, Cambridge, MA 02139, USA 
        \and Observatori de Ca l'Ou, Carrer de dalt 18, Sant Mart{\'i} Sesgueioles 08282, Barcelona, Spain 
        \and Department of Astronomy \& Astrophysics, 525 Davey Laboratory, The Pennsylvania State University, University Park, PA, 16802, USA 
        \and Center for Exoplanets and Habitable Worlds, 525 Davey Laboratory, The Pennsylvania State University, University Park, PA, 16802, USA 
        \and NASA Ames Research Center, Moffett Field, CA 94035 USA 
        \and Department of Physics and Astronomy, The University of North Carolina at Chapel Hill, Chapel Hill, NC 27599-3255, USA 
        \and NASA Exoplanet Science Institute, IPAC, California Institute of Technology, Pasadena, CA 91125 USA 
        \and Department of Aeronautics and Astronautics, MIT, 77 Massachusetts Avenue, Cambridge, MA 02139, USA 
        \and Department of Astrophysical Sciences, Princeton University, Princeton, NJ 08544, USA 
        \and SETI Institute, Mountain View, CA 94043 USA 
        }


   \date{Received ; accepted }

  \abstract
   {Different theories have been developed to explain the origins and properties of close-in giant planets, but none of them alone can explain all of the properties of the warm Jupiters (WJs, P$_{\rm orb}$ = 10 - 200 days). One of the most intriguing characteristics of WJs is that they have a wide range of orbital eccentricities, challenging our understanding of their formation and evolution.  
}
   {The investigation of these systems is crucial in order to put constraints on formation and evolution theories. \textit{TESS} is providing a significant sample of transiting WJs around stars bright enough to allow spectroscopic follow-up studies.}
   {We carried out a radial velocity (RV) follow-up study of the \textit{TESS} candidate TOI-4515\,b with the high-resolution spectrograph HARPS-N in the context of the GAPS project, the aim of which is to characterize young giant planets, and the TRES and FEROS spectrographs. We then performed a joint analysis of the HARPS-N, TRES, FEROS, and \textit{TESS} data in order to fully characterize this planetary system.}
   {We find that TOI-4515\,b orbits a 1.2 Gyr-old G-star, has an orbital period of P$_\mathrm{b}$\,=\,$15.266446 \pm 0.000013$ days, a mass of M$_\mathrm{b}$\,=\,$2.01\pm0.05$ M$_{\rm J}$, and a radius of R$_\mathrm{b}$\,=\,$1.09\pm0.04$ R$_{\rm J}$. We also find an eccentricity of e\,=\,$0.46\pm0.01$, placing  this planet among the WJs with highly eccentric orbits. As no additional companion has been detected, this high eccentricity might be the consequence of past violent scattering events.}
   {}


\titlerunning{GAPS L: TOI-4515b: an eccentric warm Jupiter orbiting a 1.2 Gyr old G-star}
\authorrunning{Carleo et al.}
   \maketitle
%

\section{Introduction}
\label{sec:intro}

Warm Jupiters (WJs) are gas giant exoplanets with orbital periods of between 10 and 200 days (e.g., \citealt{DawsonJohnson2018}), which make them challenging targets for transit detection and radial velocity (RV) follow-up studies compared to their shorter-orbit counterparts (hot Jupiters, HJs). It is for this reason that WJs have so far received less attention. However, these objects are very interesting targets for investigation in order to put constraints on and improve planetary migration theories. Indeed, the observed WJs present a wide span of values in planetary properties, especially in eccentricity \citep{Dong2021}. They have been detected in low- to moderate-eccentricity orbits (e $\lesssim$ 0.4, e.g., \citealt{Brahmetal2016,Niedzielskietal2016,Smithetal2017}), as well as in highly eccentric orbits (e.g., \citealt{Dawsonetal2012,Ortizetal2015,Guptaetal2023, DongHuang2021}). This makes it difficult to link them to a single origin or migration channel; instead, it is thought that the WJ region (around 0.1-1 AU) can be populated by different formation mechanisms (Figure 1 from \citealt{DawsonJohnson2018}). However, very recently, \citealt{Rodriguez2023} showed that a wide range of eccentricities are also present in the HJ regime, for orbital periods longer than 5 days, where tidal forces are not strong. This might suggest that the WJ regime and the longer-period HJ regime are connected. 

One explanation of the variable eccentricities of  WJs is linked to the possible dynamical coupling with a companion in the system \citep{PetrovichTremaine2016}. For example, \cite{Dongetal2014} showed that known WJs with high eccentricities (e $\gtrsim$ 0.4) tend to have a massive planetary or stellar companion in a long-period orbit (see their Figure 4). The architectures of these systems suggest that eccentric WJs might have undergone high-eccentricity migration excited by the outer companion. Furthermore, \cite{Dawsonetal2012} and \cite{Ortizetal2015} present highly eccentric WJs together with the presence of outer companions. However, there are certainly eccentric WJs without detected massive companions (e.g. \citealt{Schlecker2020}). On the other hand, WJs with no detected giant companion tend to have lower but still significant eccentricities that peak around 0.2 \citep{Dongetal2014}. However, the mechanisms used to explain the eccentricities have an impact on the stellar obliquity as well. High eccentricities might be correlated to high obliquities (e.g., \citealt{Dongetal2023}) and cool stars orbited by distant giant planets like WJs present high obliquities \citep{Albrechtetal2022}. Discovering and studying these systems 
in detail is important in order to increase the statistics and better relate the observed properties to theory.  


In this paper, we present the discovery of an eccentric WJ, TOI-4515\,b (TYC 1203-1161-1), which has no known outer companion. The target was selected for the program GAPS-Young Objects (GAPS-YO, \citealt{carleo2020,carleo2021,nardiello2022}), the aim of which is to validate, confirm, and determine the mass of transiting planets around young stars ($\lesssim$ 700 Myr). GAPS-YO is an ongoing program at the Telescopio Nazionale Galileo (TNG) using the HARPS-N spectrograph in the framework of the GAPS program \citep{Covinoetal2013}. A clear photometric modulation is seen in TESS data, suggesting a moderately young age (a few hundred million years). The subsequent analysis presented in this paper indicates that the star is likely slightly older than 1 Gyr, and is therefore more mature than the other targets considered in the GAPS-YO program. Nevertheless, we decided to complete the RV monitoring in order to determine the mass of the WJ, considering the importance of investigating these systems, both by increasing the number of this population and by fully characterizing them. 

The paper is organized as follows. The observations and available data are presented in Sect. \ref{sec:observations}; our analysis aimed at retrieving the stellar properties is described in Sect. \ref{sec:star}. The planet validation is presented in Sect. \ref{sec:planet validation}, and the joint fit with the retrieved planetary system parameters is presented in Sect. \ref{sec:Planetary system analysis}. We then present a discussion in Sect. \ref{sec:discussion} and draw conclusions in Sect.\ref{sec:conclusion}.

\section{Observations and data reduction}\label{sec:observations}

\subsection{Photometric data}

\subsubsection{TESS} \label{sub:tess}
TOI-4515 was observed in both short (120s) and long (1800s) cadence by TESS in Sectors 17 (from UT 2019 October 7 to November 2, only in long cadence mode), 42, 43 (from UT 2021 August 20 to 2021 October 12, program IDs: GO-4195, GO-4231, GO-4191), and 57 (UT 2022 September 30 to October 29, GO-5054). The Science Processing Operations Center (SPOC) conducted a transit search of Sector 42 on UT 2021 September 22 with an adaptive, noise-compensating matched filter \citep{Jenkins2002ApJ...575..493J,Jenkins2010SPIE.7740E..0DJ,Jenkins2020TPSkdph}, producing a TCE for which an initial limb-darkened transit model was fitted \citep{Li:DVmodelFit2019} and a suite of diagnostic tests were conducted to help make or break the planetary nature of the signal \citep{Twicken:DVdiagnostics2018}. The transit signature was also detected in a search of full-frame image (FFI) data by the Quick Look Pipeline (QLP) at Massachusetts Institute of Technology (MIT) \citep{Huang2020RNAAS...4..204H,huang2020RNAAS...4..206H}. The TESS Science Office (TSO) reviewed the vetting information and issued an alert on UT 2021 October 21 \citep{guerrero:TOIs2021ApJS}. The signal was repeatedly recovered as additional observations were made in sectors 42, 43, and 57, and the transit signature passed all the diagnostic tests presented in the Data Validation reports on UT 2023 February 6. The host star is located within 0.61$\pm$2.53 arcsec of the source of the transit signal.

In this work, we adopted both the short- and long-cadence light curves in order to validate the transits, study the stellar activity, and extract planetary information. Long-cadence light curves are extracted and corrected using the PATHOS pipeline described in detail in \citealt{Nardielloetal2019,Nardielloetal2020,Nardielloetal2021,Nardiello2020}. For our analysis, we cleaned the light curves excluding all the points flagged with \texttt{DQUALITY>0}.
For the short-cadence light curves, we adopted the Presearch Data Conditioning Simple Aperture Photometry (PDCSAP) light curves (\citealt{2012PASP..124.1000S,Stumpe2012,2014PASP..126..100S}). 

\begin{figure}
  \centering
  \includegraphics[width=0.45\textwidth]{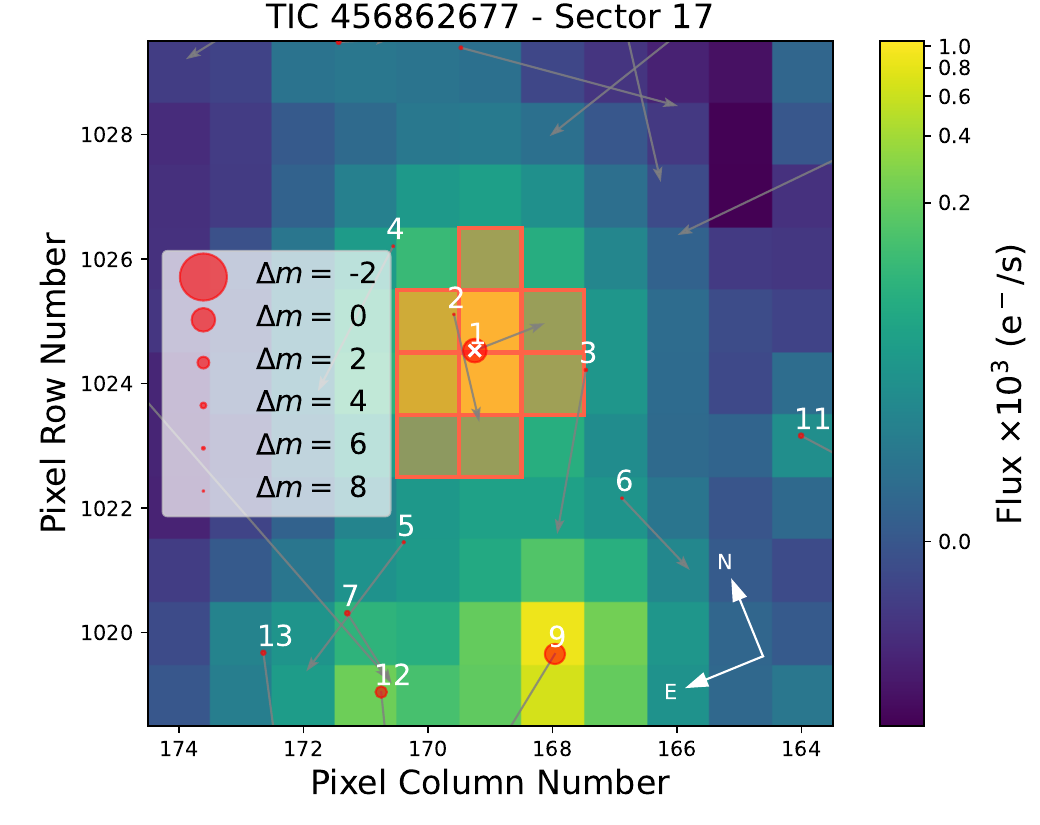}
  \caption{TESS TPF of Sector 17 for TOI-4515. The color bar indicates the electron counts for each pixel. The orange squares show the pixels selected to obtain the aperture photometry by the TESS pipeline. All the sources in Gaia DR3 are overplotted and represented with circles of different sizes according to the G-mag difference with respect to our target (see the legend); this was done with the {\tt tpfplotter} code \citep{aller2020}. Gray arrows indicate the direction of the proper motions for all the sources in the plot.
  \label{fig:tpf}}
\end{figure}

In order to verify that no additional sources could contaminate the TESS flux of TOI-4515, we inspected the Target Pixel Files (TPF), which contain the original CCD pixel observations. The code overplots ---on the TPF image--- all the sources present in the Data Release 3 of Gaia with a specific contrast magnitude with respect to our target (in this case we set $\Delta m = 8$), and highlights the aperture mask employed by the TESS pipeline to measure the SAP flux.
According to Figure \ref{fig:tpf}, only one potentially contaminating source is included within the aperture mask (Gaia DR3 ID 289464440515394944), and has a Gmag equal to 19.67 ($\Delta m \sim 7.9$). We note that the crowding reported for each sector in which TOI-4515 is observed is always less than 1\% (based on an analysis of the TIC-8 catalog and the pixel response functions reconstructed from dithered data sets obtained at the beginning of the mission). This level of contamination is properly corrected in the PDCSAP curve by the SPOC pipeline.
In accordance with the approach outlined in \cite{2022MNRAS.516.4432M}, we used \textit{Gaia} DR3 data to detect nearby contaminating stars that might be blended eclipsing binaries (BEBs) and measure the dilution factor, which denotes the total flux from contaminant stars that fall into the photometric aperture divided by the flux contribution of the target star. Our analysis reveals that, besides TOI-4515, none of the \textit{Gaia} resolved stars within a radius of ten TESS pixels from the target star can reproduce the transit signal of TOI-4515.01. Additionally, our investigation indicates an almost negligible dilution factor of 0.01. 

\subsubsection{WASP archival data} \label{sub:wasp}
The field of TOI-4515 was observed by the WASP transit-search survey \citep{2006PASP..118.1407P} between 2004 and 2014, accumulating a total of 90\,000 photometric data points in a broad, visual passband.  Observations on each clear night spanned up to 150 days in each year, observing the field with a $\sim$\,15 min cadence. TOI-4515 is the only bright star within the 48 arcsec photometric extraction aperture. Looking at the WASP data, we find two possible transit features that match up with the ephemeris in Sect. \ref{tab:fit_params}. While these might be real pre-detections of the transit, this is not certain.

\subsubsection{KeplerCam}

We observed a full transit of TOI-4515.01 on UT 2021 October 25 from KeplerCam on the 1.2\,m telescope at the Fred Lawrence Whipple Observatory using a Sloan $i'$ band filter. The $4096\times4096$ Fairchild CCD 486 detector has an image scale of $0\farcs672$ per $2\times2$ binned pixel, resulting in a $23\farcm1\times23\farcm1$ field of view. The image data were calibrated and photometric data were extracted using {\tt AstroImageJ} \citep{Collins:2017}. We used circular photometric apertures with radius $4\arcsec$ centered on TOI-4515. The target star aperture excluded flux from the nearest known neighbor in the Gaia DR3 and TICv8 catalogs (TIC 620491915), which is $\sim14\arcsec$ north of TOI-4515. The target star light curve was linearly detrended using the full width at half maximum (FWHM) of the target star point spread function in each image. A clear transit-like event was detected and the light curve included in the global model described in Section \ref{sec:Planetary system analysis}.

\subsubsection{CALOU}

TOI-4515.01 was first released as a TESS Object of Interest from a Sector 42 SPOC Data Validation Report. While the period was released as $\sim15.265$ days, a TESS data gap that occurred due to the spacecraft data download phase of operation allowed a potential planet candidate orbital period alias of 7.6325 days if an additional transit occurred during the data gap. We therefore observed a predicted full transit window ---assuming the 7.6325 day period--- to check for a transit-like event on an epoch that corresponds to the TESS data gap. We observed the would-be transit window using the Observatori de Ca l'Ou (CALOU), a private observatory in Sant Martí Sesgueioles, near Barcelona Spain, in the Rc passband on UT 2022 January 1. The 0.4\,m telescope is equipped with a $1024\times1024$ pixel FLI PL1001 camera with an image scale of $1\farcs$14 pixel$^{-1}$, resulting in a $21\arcmin\times21\arcmin$ field of view. The images were calibrated and differential photometric data were extracted using {\tt AstroImageJ}. We used a circular photometric aperture with a radius of $10\arcsec$ centered on TOI-4515 and ruled out the expected $\sim16$\,ppt deep event, confirming the true period to be $\sim15.265$ days.

\subsubsection{LCOGT}

We observed a full transit window (at the $\sim15.265$\,d ephemeris) of TOI-4515.01 on 2022 November 11 in Sloan $g'$ band using the Las Cumbres Observatory Global Telescope \citep[LCOGT;][]{Brown:2013} 0.4\,m network node at Cerro Tololo Inter-American Observatory (CTIO). The 0.4\,m telescopes are equipped with $2048\times3072$ pixel SBIG STX6303 cameras with an image scale of 0$\farcs$57 pixel$^{-1}$, resulting in a $19\arcmin\times29\arcmin$ field of view. The images were calibrated by the standard LCOGT {\tt BANZAI} pipeline \citep{McCully:2018} and differential photometric data were extracted using {\tt AstroImageJ}. We used circular photometric apertures with a radius of $5\farcs7$ centered on TOI-4515. The target star aperture excluded flux from the nearest known neighbor in the Gaia DR3 and TICv8 catalogs (TIC 620491915), which is $\sim14\arcsec$ north of TOI-4515. A clear transit-like event was detected and the light-curve data are included in the global model described in Section \ref{sec:Planetary system analysis}.

\subsection{Spectroscopic data}

\subsubsection{HARPS-N}
Within the GAPS Project, and in particular the subprogram focused on the Young-Objects follow-up \citep{carleo2020}, we observed TOI-4515 with the high-resolution spectrograph HARPS-N \citep{Cosentinoetal2014} mounted on the TNG in La Palma, Spain. The 25 observations span a period of time between UT 2021 December 12 and UT 2022 November 9, with an exposure time of 1800s and an average [min, max] signal-to-noise ratio (S/N) of 36 [18, 49]. The data were reduced with the offline version of HARPS-N data reduction software (DRS) through the Yabi web application (\citealt{yabi}) installed at IA2 Data Center\footnote{\url{https://www.ia2.inaf.it}}. The RV measurements were obtained using a G2 mask template and a cross-correlation function (CCF) width of 40\,km\,s$^{-1}$, with an average precision of 3\,m\,s$^{-1}$. The list of RVs is presented in Table \ref{tab:rvdata}, together with the chromospheric activity index \logrhk (see Sec. \ref{sec:activity}).

\subsubsection{TRES}
We observed TOI-4515 a total of 18 times between UT 2021 October 29 and UT 2022 September 27 using the Tillinghast Reflector Echelle Spectrograph \citep[TRES;][]{fureszetal2008}\footnote{\url{http://www.sao.arizona.edu/html/FLWO/60/TRES/GABORthesis.pdf}\citep{gaborthesis}} on the 1.5m Tillinghast Reflector in order to measure the RV orbit of the planet and precisely constrain key parameters, such as the orbital eccentricity and mass of the companion. Attached to the 1.5m telescope Fred L. Whipple Observatory (FLWO) on Mt. Hopkins, AZ. TRES has a spectral resolution of 44,000, and for TOI-4515, we obtained a typical S/N per resolution element of $\sim$30. The data were reduced and the RVs were extracted following the techniques of \citet{Buchhave:2010} and \citet{Quinn:2012}; these are reported in Table \ref{tab:rvdata}.

\subsubsection{FEROS}

We observed TOI-4515 with the FEROS spectrograph \citep{Kaufer99} at the MPG 2.2m telescope at La Silla (resolving power $\mathrm{R=50\,000}$) in the context of the Warm gIaNts with tEss (WINE) collaboration. Five spectra were obtained between UT 2021 November  26 and UT 2022 October 18 in Object-Calibration mode, with an exposure time of 900s, under Program IDs 0108.A-9003, 0109.A-9003, and 0110.A-9011. The spectra were reduced with the \texttt{ceres} pipeline \citep{Brahm17CERES}. The list of RVs is presented in Table \ref{tab:rvdata}. 

\subsection{High-contrast imaging}
\label{sec:imaging}

\subsubsection{SOAR}
\label{sec:soar}

\begin{figure}
    \centering
    \includegraphics[width=\linewidth]{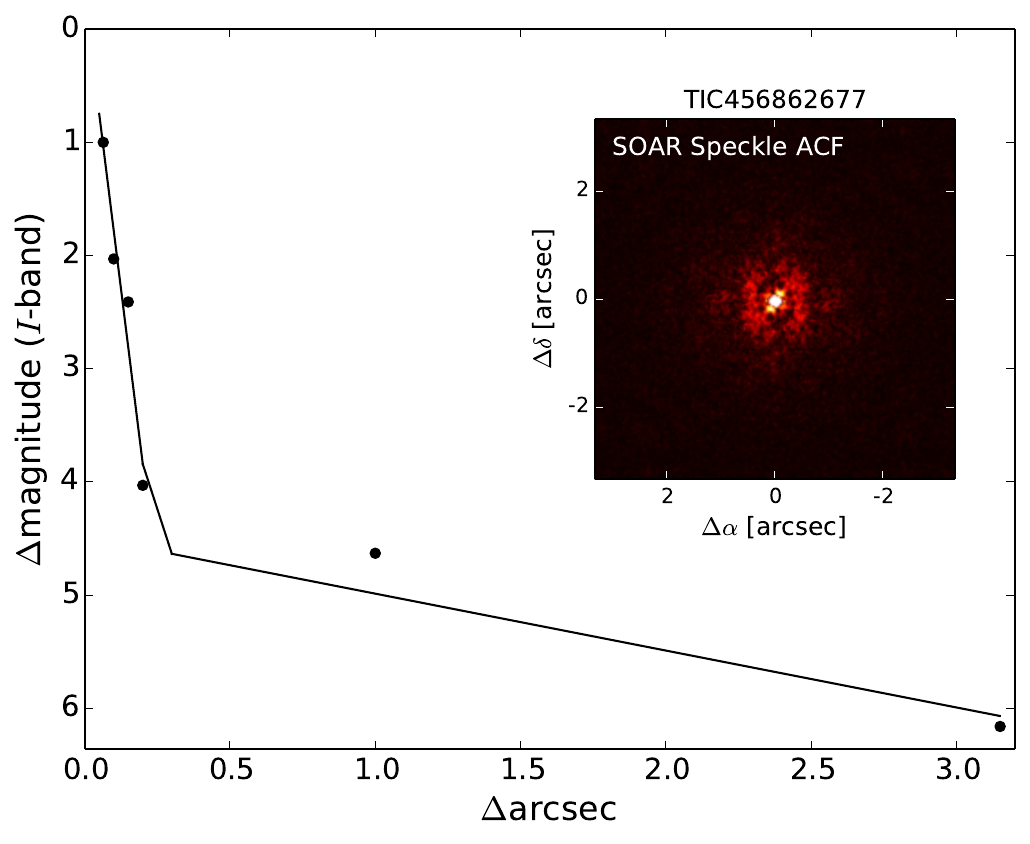}
    \caption{ 5$\sigma$ sensitivity limits and inset speckle auto-correlation function for the SOAR observation of TOI-4515 taken in the visible I-band. No nearby sources were detected within 3\arcsec of the target star.}
    \label{fig:soar}
\end{figure}

High-angular resolution imaging is needed to search for nearby sources that can contaminate the TESS photometry, resulting in an underestimated planetary radius, or be the source of astrophysical false positives, such as background eclipsing binaries. We searched for stellar companions to TOI-4515 with speckle imaging on the 4.1m Southern Astrophysical Research (SOAR) telescope \citep{2018tokovinin} on UT 2021 November 20 , observing in Cousins I-band, a similar visible bandpass to that of TESS. This observation was sensitive to a 
star that is 6.2 magnitudes fainter found at an angular distance of 1 arcsec from the target. More details of the observations within the SOAR TESS survey are available in \citet{2020ziegler}. The 5$\sigma$ detection sensitivity and speckle auto-correlation functions from the observations are shown in Figure \ref{fig:soar}. No nearby stars were detected within 3\arcsec of TOI-4515 in the SOAR observations.

\subsubsection{NESSI}
\label{sec:nessi}

\begin{figure}
    \centering
    \includegraphics[width=\linewidth]{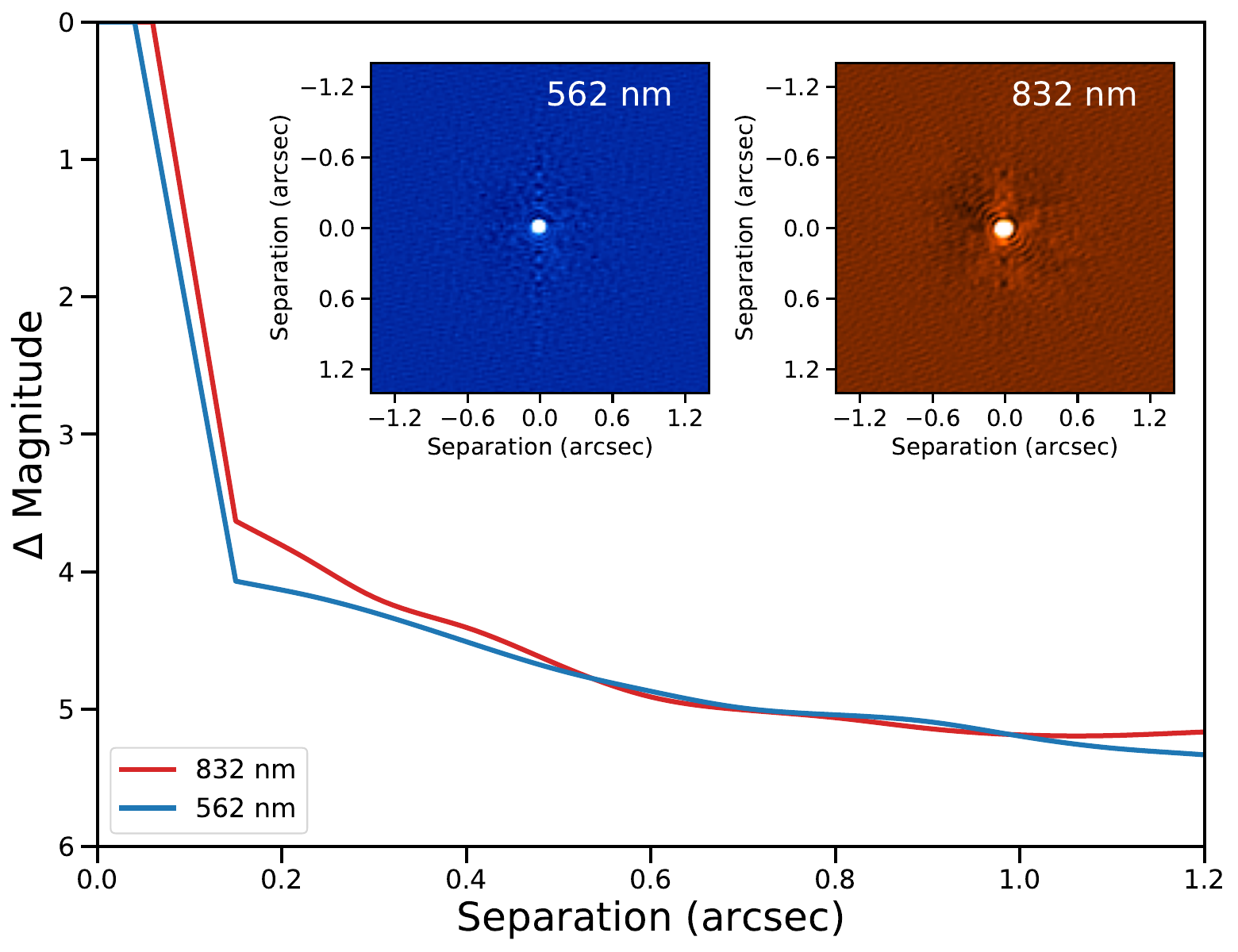}
    \caption{Reconstructed NESSI speckle images and $5\sigma$ contrast limits from simultaneous diffraction-limited exposure sequences using the 562 nm filter on the blue camera and the 832 nm filter on the red camera. No nearby sources are detected.}
    \label{fig:nessi}
\end{figure}

We place further constraints on the presence of nearby sources that could contaminate the photometry or produce a false positive using the NN-explore Exoplanet Stellar Speckle Imager \citep[NESSI;][]{Scott2018} on the WIYN 3.5m telescope at Kitt Peak National Observatory. We observed TOI-4515 with NESSI on the night of UT 2021 October 29; simultaneous 1 min sequences of 40 ms diffraction-limited exposures were taken in the 562 nm and 832 nm filters on the blue and red NESSI cameras, respectively. The reconstructed speckle images generated following the methods described by \citet{Howell2011} are shown alongside $5\sigma$ contrast curves in Figure \ref{fig:nessi}. The NESSI data rule out the presence of nearby stellar companions and background sources down to $\Delta$mag$\approx4$ at a separation of 0.2'' and $\Delta$mag$\approx5.2$ at a separation of 1''.

\section{Stellar analysis}\label{sec:star}

The methods for stellar characterization follow the approach of our previous investigations
\citep[e.g., ][]{nardiello2022}; we consider a variety of complementary methods for age determination
\citep{desidera2015}
and exploit the available high-resolution spectra \citep{2020baratella} from which we obtained a co-added spectrum with high S/N.
TOI-4515 has not been the subject of a targeted study until now.
The stellar parameters adopted from the literature or derived below are summarized in Table \ref{t:star_param}.
The constraints on the presence of additional companions, both planetary and stellar, over the full separation range are presented in Sect. \ref{sec:additionalcompanions}.

For the derivation of the photometric temperatures, we exploited the adopted (unreddened) magnitudes and colors from Table \ref{t:star_param}, the tables by \citet{pecaut2013}
\footnote{\url{https://www.pas.rochester.edu/~emamajek/EEM_dwarf_UBVIJHK_colors_Teff.txt}, version 2022.04.16 }, hereafter referred to as the Mamajek tables, and the reddening derived through maps in the PLATO Input Catalog \citep[PIC, ][]{pic}, which amounts to E(B-V)=0.027$\pm$0.018. The resulting photometric $T_{\rm eff}$ is 5419$\pm$100 K.

\subsection{Stellar parameters and iron abundance}
\label{sec:spec_anal}

We derived stellar parameters with the standard equivalent width (EW) method by analyzing the HARPS-N co-added spectrum. First, we estimated the input effective temperature \teff  using the calibrated relations by \cite{2010casagrande} ($V$-$K_s$ and $J$-$K_s$ de-reddened color indexes) and the relations by \cite{2021mucciarelli} ($G$-$K_s$, $G_{\rm BP}$-$K_s$, and $G_{\rm BP}$-$G_{\rm RP}$ de-reddened color indexes). We adopted E(B-V)=0.027 to correct the color indexes (this value was taken from the TESS Input Catalog version 8.2 - TIC v8.2, \citealt{TICv82}). This temperature estimate was then used to derive the initial guess for the surface gravity \logg$_{\rm{trig}}$ from the classical equation exploiting the Gaia parallax, adopting the mass from the TIC v8.2, of namely M=0.944 M$_{\sun}$. Finally, we derived the microturbulence velocity parameter $\xi$ from the relation by \cite{2016dutraferreira}. The input values are \teff($V$-$K_s$)=5426 K, \teff($J$-$K_s$)=5342 K, \teff($G_{\rm BP}$-$G_{\rm RP}$)=5401 K, \teff($G$-$K_s$)=5414 K, \teff($G_{\rm BP}$-$K_s$)=5405 K, \logg$_{\rm{trig}}=4.57\pm0.07$ dex, and $\xi=0.89\pm0.07$ \kms. 

The line list adopted is from \cite{2020baratella}: from this, we measured EWs of the iron lines using the code ARES v2 \citep{2015sousa}. We discarded lines with errors larger than 10$\%$ and with EW$>120$\,m\AA\, in order to avoid issues with the Guassian fit of the line performed by ARES. We used the code q2 \citep{2014ramirez}, which is based on the 2019 MOOG version (\citealt{sneden1973}), and the ATLAS9 1D LTE model atmospheres, with new opacities (ODFNEW; \citealt{castellikurucz2003}), in order to derive the spectroscopic photospheric parameters. 
The final solution of our analysis is \teff=$5447\pm29$\,K, \logg$=4.48\pm0.10$ dex, $\xi=1.06\pm0.09$\,\kms , and [Fe/H]=$0.05\pm0.03$ considering a solar A(Fe)=7.49 obtained from the analysis of a HARPS-N spectrum and using the same line list (see \citealt{2020baratella}). The error on \teff is the internal error coming from the code q2, while the typical systematic error in the spectroscopic temperature is considered to be 100 K.

We obtained an independent estimate of the host star parameters from the TRES spectra. We used the Stellar Parameter Classification (SPC) package \citep{Buchhave:2012, buchhave2014}, estimating [m/H]=0.10$\pm$0.08 dex, $T_{\rm eff}$=5487$\pm$50 K, a sky-projected rotational velocity of 3.7$\pm$0.5 km\,s$^{-1}$ (not corrected for macroturbulence), and $\log g$=4.57$\pm$0.10. We find good agreement at 1 sigma between the values obtained from the HARPS-N and the TRES spectra

\subsection{Spectral energy distribution}
\label{sec:sed}

\begin{figure}
    \centering
    \includegraphics[width=0.75\linewidth,trim=75 80 80 90,clip,angle=90]{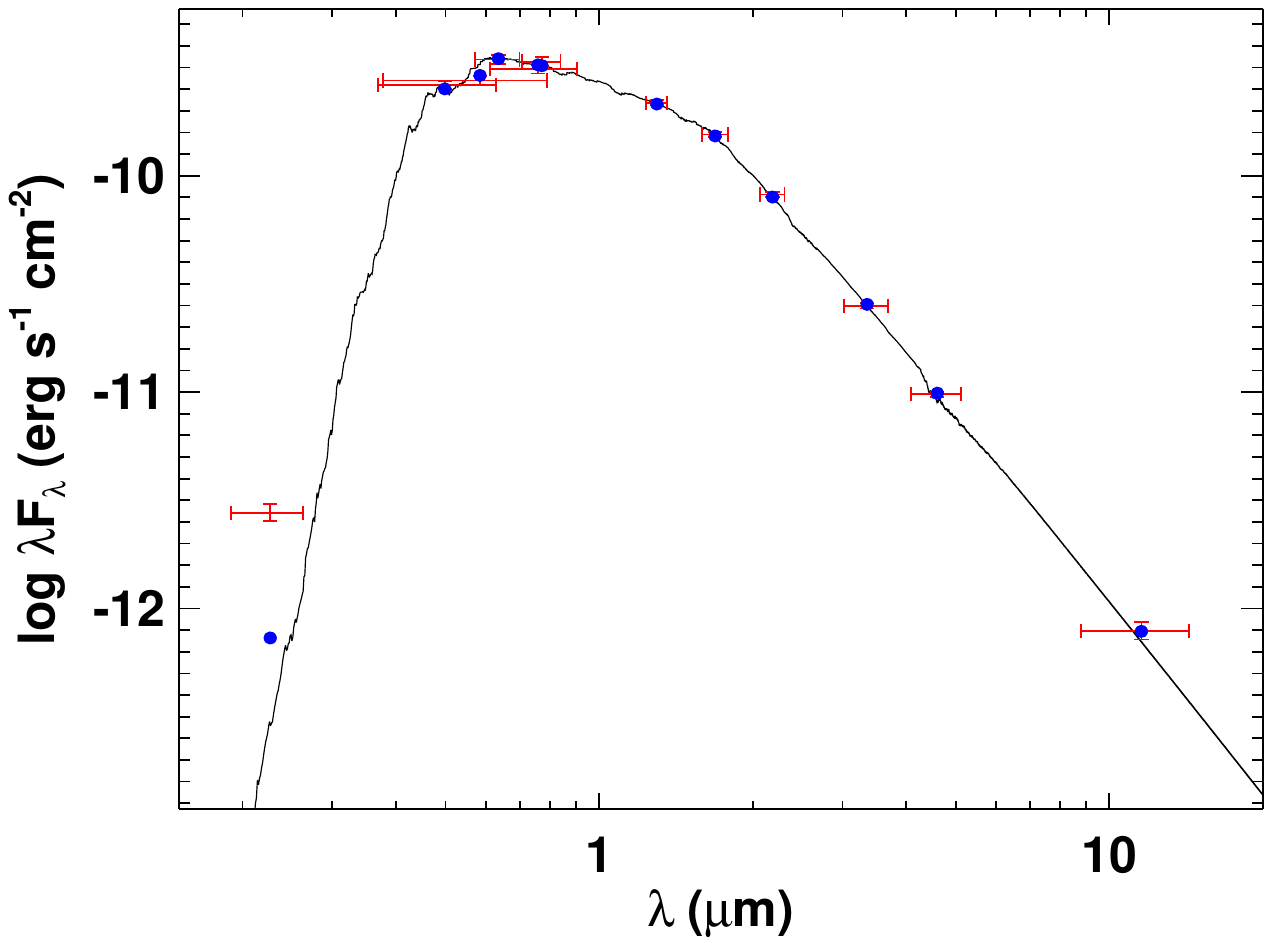}
    \caption{Spectral energy distribution of TOI-4515. Red bars represent the observed photometric measurements, and the horizontal bars represent the effective width of the passband. Blue symbols are the model fluxes from the best-fit NextGen atmosphere model (black).}
    \label{fig:sed}
\end{figure}

As an independent determination of the basic stellar parameters, we performed an analysis of the broadband spectral energy distribution (SED) of the star together with the {\it Gaia\/} DR3 parallax \citep[with no systematic offset applied; see, e.g.,][]{StassunTorres:2021} in order to determine an empirical measurement of the stellar luminosity and radius, following the procedures described in \citet{Stassun:2016,Stassun:2017,Stassun:2018}. We obtained the $gri$ magnitudes from {\it APASS}, the $JHK_S$ magnitudes from {\it 2MASS}, the W1--W3 magnitudes from {\it WISE}, the $G_{\rm BP} G_{\rm RP}$ magnitudes from {\it Gaia}, and the near-ultraviolet (NUV) magnitude from {\it GALEX}. Together, the available photometry spans the full stellar SED over the wavelength range 0.2--22~$\mu$m (see Figure \ref{fig:sed}).

We performed a fit using NextGen stellar atmosphere models, with the free parameters being the $T_{\rm eff}$ and the extinction $A_V$, which we limited to the maximum line-of-sight value from the Galactic dust maps of \citet{Schlegel:1998}. We also adopted the metallicity determined from the spectroscopic analysis above. The resulting fit (Figure \ref{fig:sed}) has a reduced $\chi^2$ of 0.9, excluding the {\it GALEX} NUV flux, which indicates a moderate level of activity \citep{Findeisen:2011}, with $A_V = 0.03 \pm 0.03$ and $T_{\rm eff} = 5335 \pm 75$~K. Integrating the (unreddened) model SED gives the bolometric flux at Earth, $F_{\rm bol} = 4.760 \pm 0.055 \times 10^{-10}$ erg~s$^{-1}$~cm$^{-2}$. Taking the $F_{\rm bol}$ together with the {\it Gaia\/} parallax directly gives the bolometric luminosity, $L_{\rm bol} = 0.5573 \pm 0.0068$~L$_\odot$, which with the $T_{\rm eff}$ 
gives the stellar radius, $R_\star = 0.875 \pm 0.025$~R$_\odot$. In addition, as a consistency check, we can estimate the stellar mass from the 
$R_\star$ together with the spectroscopically determined $\log g$, giving $M_\star = 1.04 \pm 0.10$~M$_\odot$.

\subsection{Projected rotational velocity}
\label{sec:rot_velo}
From $T_{\rm eff}$, $\log g$, $\xi$, and [\ion{Fe}/H] fixed to the final values found in Sect.\,\ref{sec:spec_anal}, we measured the stellar projected rotational velocity ($v\sin{i_{\star}}$) using the same MOOG code as above and applying the spectral synthesis of three regions around 5400, 6200\,, and 6700\,\AA. We adopted the same grid of model atmospheres as in Sect.\,\ref{sec:spec_anal} and, after fixing the macroturbulence velocity to the value of 2.5\,km\,s$^{-1}$ from the relationship by \cite{Breweretal2016}, we find a $v\sin{i_{\star}}$ of $3.4\pm0.5$\,km\,s$^{-1}$. This value is compatible with the one calculated through Equation 7 in \cite{Raineretal2023}, of namely $3.6\pm0.5$\,km\,s$^{-1}$.

\subsection{Lithium abundance}
\label{sec:lithium}

We also derived the lithium abundance $A_{\rm Li}$ from the measured lithium EW ($< 2.1$\,m\AA) and considering our stellar parameters previously derived together with the nonlocal thermodynamic equilibrium (NLTE) corrections by \cite{Lindetal2009}. We could only obtain an upper limit of $<0.5$\,dex on the lithium abundance. We obtain the same value also considering the synthesis analysis based on the MOOG code, after fixing the stellar parameters to those derived in Sects.\,\ref{sec:spec_anal} and \ref{sec:rot_velo}. This upper limit supports an age of greater than that of the Hyades.

\begin{figure}
\includegraphics[width=0.45\textwidth]{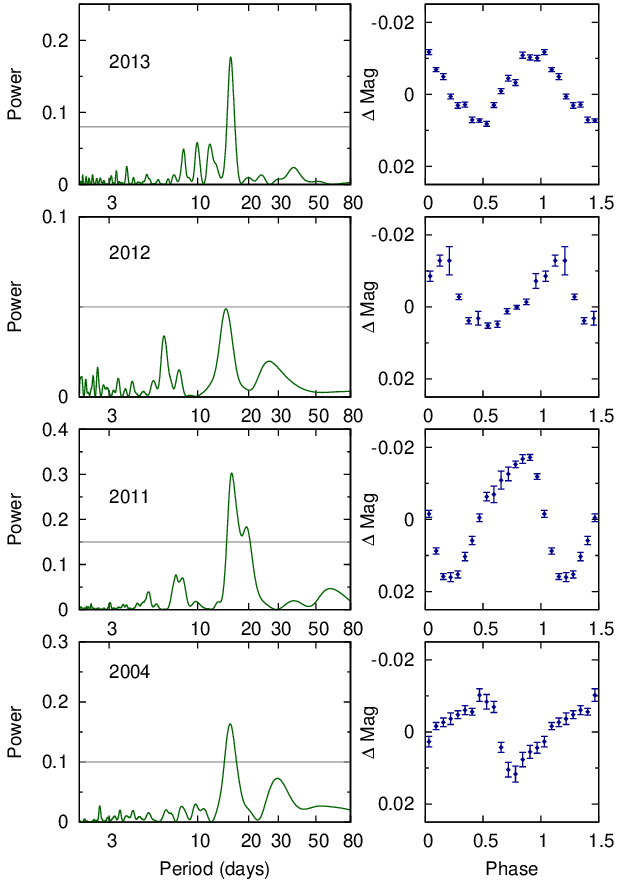}
  \caption{Periodograms of the SuperWASP data for TOI-4515 in selected years (left) along with folds of the data on the 15.6 d rotational period (right). The horizontal line in the periodograms is the estimated 1\% likelihood false-alarm level.}
\label{fig:wasp}
\end{figure}

\subsection{Rotation period}
\label{sec:rotation}

We searched each year of the SuperWASP photometric data for a rotational modulation using methods outlined in \citet{2011PASP..123..547M}.  There is a persistent and highly significant periodicity with a period of 15.6 $\pm$ 0.3 days (see Figure\,\ref{fig:wasp}), where the error makes some allowance for phase shifts caused by changing star-spot patterns.  The amplitude changes between years, varying between 5 mmag and 16 mmag.

\begin{figure}
  \centering
  \includegraphics[trim=0 0 0 50, angle=90,width=0.5\textwidth]{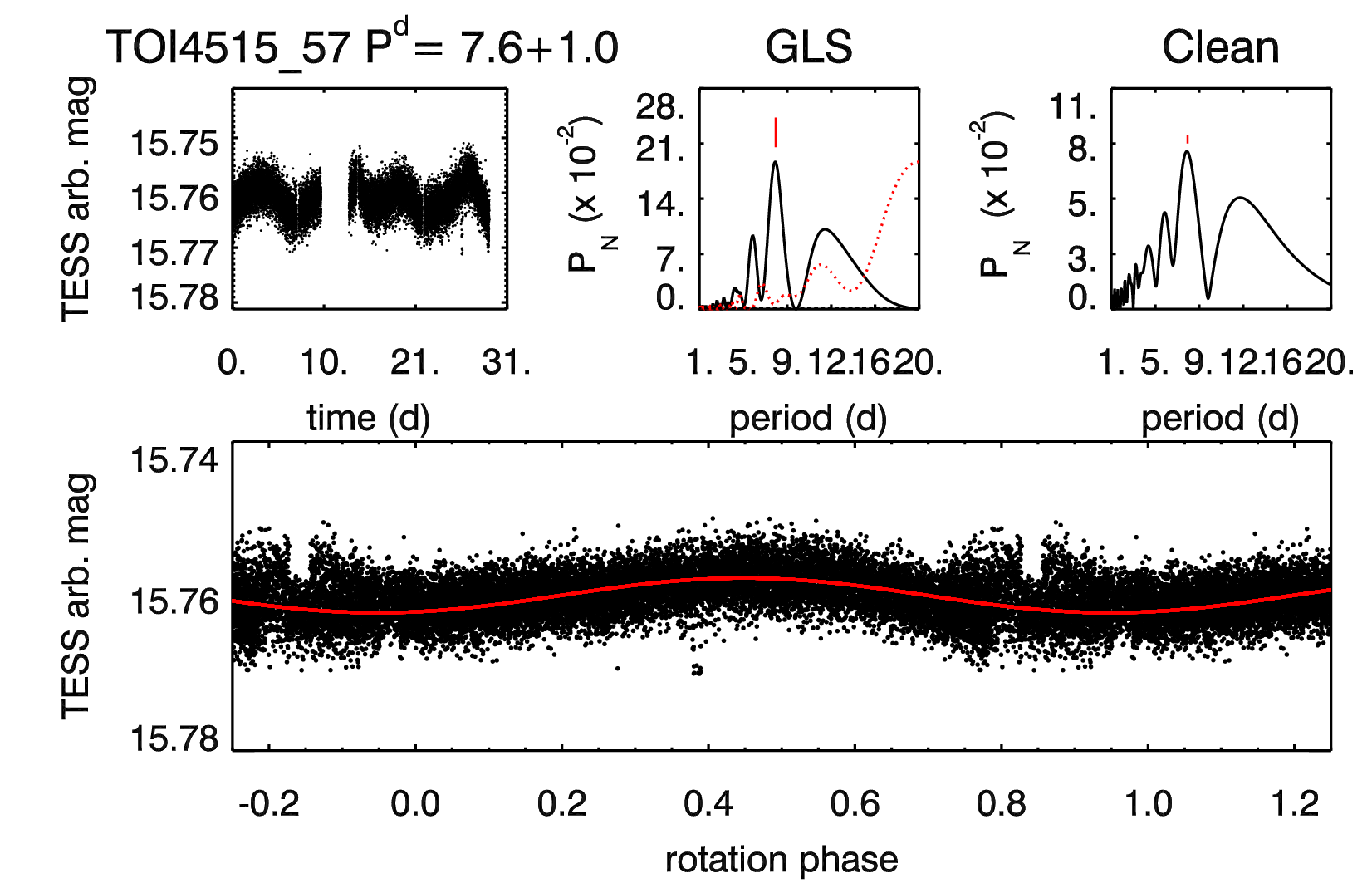}
  \caption{Results of the periodogram analysis of the TESS photometry (Sector 57) of TOI-4515. Top-left panel: Photometric time series.  Red vertical lines indicate the epochs of primary and secondary minimum.  Top-middle panel: GLS periodogram (solid black line) with the window function  overplotted (dotted red line). Top-right panel: CLEAN periodogram. Bottom panel:  TESS light curve phased with the primary periodogram periodicity, that is with the 7.6 d half stellar rotation period.  \label{fig:TOI4545_lc}}
\end{figure}

We analyzed the photometric data time series from the four TESS Sectors separately and in a single combined series using the Generalized Lomb-Scargle (GLS) and Clean periodograms, following the procedure described in \citet{messina2022}.\\
A highly significant periodicity is found at P = 7.64$\pm$0.03\,d in the combined series and similar  periodicities, within the uncertainties, are detected in each single sector: P = 7.9$\pm$1.2\,d in Sector\,17, P = 7.4$\pm$0.5\,d in the consecutive Sectors\,45 and 46, P = 7.6$\pm$1.0\,d in Sector\,57. As an example, a summary of the results from the periodogram analysis for Sect.\,57 are reported in Figure\,\ref{fig:TOI4545_lc}. This periodicity is most likely linked to the rotation period of the star. As this periodicity is about half the rotation period derived from SuperWASP data (P = 15.6\,d), we considered the possibility
that the true period is twice the observed one. This circumstance may occur when dominant star spots of similar size occur on the stellar surface at longitude separated by about 180 deg, leading to similar features in the light curve on two occasions during one rotation period (double-dip variables). One example is represented by TOI-1807 \citep{nardiello2022} (for a comprehensive discussion on period-doubling effects, see e.g., \citealt{CollierCameron2009}).  
Whereas in the Sectors 17, 45, and 46, a gap in the data series prevented detection of the double dip, in the last Sector 57 it is quite clearly visible. More specifically, we see two consecutive minima of different depths separated by 7.6\,d (marked by red ticks in Figure\,\ref{fig:TOI4545_lc}), which supports our hypothesis of a true rotation period double as inferred from the periodogram analysis, that is P = 15.26$\pm$0.06\,d. This independent estimate from TESS data is in fair agreement with that from SuperWASP and with the rotation period expected on the basis of the rotation--chromospheric activity relation by \citet{mamajek2008} (see Sect.\,\ref{sec:activity}). This rotation period also implies an age in agreement with the other age indicators analyzed in the present investigation, such as lithium (Sect. \ref{sec:lithium}), chromospheric
activity (Sect. \ref{sec:activity}), and kinematics (Sect. \ref{sec:kinematics}).

We also investigated the periodogram of the HARPS-N RV time-series, as well as several activity indicators (\logrhk, S-index, Bisector, CCF Contrast, CCF FWHM, chromospheric index CRX, differential line width dLw, H-alpha, and the sodium lines Na$_1$ and Na$_2$). Only the RV time series presents a peak at P$\sim$7.5d, albeit not significant, because the RV periodogram is dominated by the planetary signal.

\subsection{Chromospheric activity}
\label{sec:activity}

Ca II H\&K emission was measured on HARPS-N spectra by exploiting the YABI tool, which is based on prescriptions by \citet{lovis2011}.
The median value 
of the S index, calibrated into the Mount Wilson scale, is 0.357$\pm$0.011.
The corresponding \logrhk \ is -4.67$\pm$0.02. 
This value is at the lower edge of Hyades members of similar color.
The expected rotation period using the \citet{mamajek2008} calibration is 13.4d, which is very close to the rotation period measured in Sect. \ref{sec:rotation} and corresponds to an age of 1.0 Gyr. No X-ray observations are available for the target.

\subsection{Kinematics}
\label{sec:kinematics}

The space velocities U,V, and W, derived following the prescriptions by \citet{uvw}, are listed in Table \ref{t:star_param}. These put the star slightly outside the kinematic locus of young stars defined by \citet{Eggen1984}, indicating an age of more than about 500 Myr and likely younger than the Sun.

\subsection{Age}
\label{sec:age}
The stellar rotation period P = 15.5$\pm$0.3\,d obtained by combining the TESS and SuperWASP results allows a gyro-chronological age estimation of about 1.2$\pm$0.2\,Gyr according to Eqs. 12--14 in \citet{mamajek2008}, with the uncertainty derived from the coefficient error propagation.
This estimate is in good agreement with those from the independent methods applied by us; that is $\ga$ 0.6\,Gyr from lithium, 1\,Gyr from chromospheric activity, and $\ga$ 0.5\,Gyr from kinematics.
We therefore adopt a stellar age of 1.2$\pm$0.2\,Gyr.

\subsection{Mass, radius, and system orientation}
\label{sec:stellarmass}
Isochrone fitting is inconclusive for age determination, as expected for an unevolved late G dwarf.
The PARAM Bayesian Interface of PARSEC models\footnote{\url{http://stev.oapd.inaf.it/cgi-bin/param_1.3}} \citep{param} yields 3.3$\pm$3.2 Gyr.
We used the same code to derive the stellar mass, allowing only the age range derived from indirect methods
as in \citet{desidera2015} and adopting the average between photometric and spectroscopic \teff. The resulting stellar mass is 0.949$\pm$0.020~M$_\odot$ (with the error being that provided by PARAM, systematic uncertainties in stellar models not included).
The stellar radius was derived using the Stefan-Boltzmann law, as in \citet{carleo2021}, adopting
the bolometric corrections given in the Mamajek tables to infer the stellar luminosity.
The resulting stellar radius is 0.860$\pm$0.030~R$_\odot$, formally  slightly smaller than but in agreement to better than 1 $\sigma$ with that derived from SED fitting (Sect. \ref{sec:sed}) and those adopted in TIC, PIC, and Gaia.

We derived a constraint on the stellar inclination angle by combining the measurements of the stellar radius, projected rotation velocity, and rotation period.  For this purpose, we adopted R$_{\star}$ = 0.86 $\pm$ 0.03 R$_\odot$, $v\sin{i_{\star}}$ = 3.4 $\pm$ 0.5\,km\,s$^{-1}$, and P$_{\rm rot}$ = 15.3 $\pm$ 1.5 days.  Although the photometric periodicity was determined with higher precision (see Sect. \ref{sec:rotation}), here we enlarged the uncertainty to 10\% to account for systematic effects such as differential rotation.  With these values, $v$ = 2$\pi$R/P$_{\rm rot}$ = 2.8 $\pm$ 0.3\,km\,s$^{-1}$, which is compatible with the measurement of $v\sin{i_{\star}}$.  Therefore, the inclination is consistent with 90 deg.  To determine the lower limit on the inclination, we used the MCMC method advocated by \cite{MasudaWinn2020}, which gave 2$\sigma$ limits of $\cos{i_{\star}}$ < 0.59 and an inclination of > 54 deg.  Therefore, the maximum difference in inclination angles between the star and planetary orbit is approximately 36 deg.

\begin{table}[!htb]
   \caption[]{Stellar properties of TOI-4515}
     \label{t:star_param}
     \small
     \centering
       \begin{tabular}{lcc}
         \hline
         \noalign{\smallskip}
         Parameter   &  \object{TOI-4515} & Ref  \\
         \noalign{\smallskip}
         \hline
         \noalign{\smallskip}
$\alpha$ (J2000)          &   01 24 44.69       & Gaia DR3    \\
$\delta$ (J2000)          &  +21 30 46.98   & Gaia DR3  \\
$\mu_{\alpha}$ (mas/yr)  &    -4.285$\pm$0.024  & Gaia DR3  \\
$\mu_{\delta}$ (mas/yr)  &    3.462$\pm$0.014  & Gaia DR3  \\
RV     (km\,s$^{-1}$)            &    12.75$\pm$0.44   & Gaia DR3   \\
$\pi$  (mas)             &    5.160$\pm$0.019 & Gaia DR3  \\
$U$   (km\,s$^{-1}$)             &      -4.63$\pm$0.23        & This paper (Sect. \ref{sec:kinematics}) \\
$V$   (km\,s$^{-1}$)             &      11.21$\pm$0.24        & This paper (Sect. \ref{sec:kinematics}) \\
$W$   (km\,s$^{-1}$)             &      -6.41$\pm$0.28       & This paper (Sect. \ref{sec:kinematics})  \\
\noalign{\medskip}
$V$ (mag)                  &    12.00$\pm$0.03     & APASS DR9    \\
$B$-$V$ (mag)                &    0.787$\pm$0.033  &  APASS DR9 \\
$G$ (mag)                  &    11.8121$\pm$0.0006  & Gaia DR2  \\
$G_{\rm BP}$-$G_{\rm RP}$ (mag)              &    0.9817              & Gaia DR2  \\
TESS (mag)              &   11.302$\pm$0.0062 & \\
J$_{\rm 2MASS}$ (mag)    &   10.625$\pm$0.027  & 2MASS  \\
H$_{\rm 2MASS}$ (mag)    &   10.190$\pm$0.030  & 2MASS  \\
K$_{\rm 2MASS}$ (mag)    &   10.134$\pm$0.018  & 2MASS  \\
\noalign{\medskip}
Spectral Type            &  G8/G9  & This paper (Sect. \ref{sec:spec_anal}) \\
$T_{\rm eff}$ (K)        &  5447$\pm$29       & This paper (spec, Sect. \ref{sec:spec_anal}) \\  
$T_{\rm eff}$ (K)        &  5487$\pm$50       & This paper (SPC/TRES,Sect.\ref{sec:spec_anal}) \\  
$T_{\rm eff}$ (K)        &  5419$\pm$100      & This paper (phot, Sect. \ref{sec:star}) \\  
$T_{\rm eff}$ (K)        &  5335 $\pm$ 75      & This paper (SED, Sect. \ref{sec:sed}) \\  
$T_{\rm eff}$ (K)        &  5433$\pm$70       & This paper (adopted, Sect. \ref{sec:stellarmass}) \\  
$\log g$                 &  4.48$\pm$0.10     & This paper (Sect. \ref{sec:spec_anal}) \\ 
${\rm [Fe/H]}$ (dex)     &  0.05$\pm$0.03     & This paper (Sect. \ref{sec:spec_anal}) \\ 
${\rm [Fe/H]}$ (dex)     &  0.10$\pm$0.08     & This paper (SPC/TRES,Sect.\ref{sec:spec_anal}) \\ 
E(B-V)                   &  0.027$\pm$0.018     & PIC \citep{pic} \\ 
\noalign{\medskip}
$S_{\rm MW}$             &   0.357$\pm$0.011    & This paper (Sect. \ref{sec:activity})\\
$\log R^{'}_{\rm HK}$    &     --4.67$\pm$0.02 &  This paper (Sect. \ref{sec:activity}) \\  
$v\sin{i_{\star}}$ (km\,s$^{-1}$)      &    3.4$\pm$0.5   & This paper (Sect. \ref{sec:rot_velo}) \\ 
$v\sin{i_{\star}}$ (km\,s$^{-1}$)      &    3.6$\pm$0.5   &  Eq. 7 in \cite{Raineretal2023} \\ 
${\rm P_{\rm rot}}$ (d)  &   15.5$\pm$0.3  &   This paper (Sect. \ref{sec:rotation}) \\
$EW_{\rm Li}$ (m\AA)     &     $<$2.1 &  This paper (Sect. \ref{sec:lithium})  \\
$A_{\rm Li}$ (dex)             &  $<$0.50 &  This paper  (Sect. \ref{sec:lithium}) \\
\noalign{\medskip}
Mass (M$_{\odot}$)       &    0.949$\pm$0.020    & This paper (adopted, Sect. \ref{sec:stellarmass}) \\
Radius (R$_{\odot}$)     &    0.860$\pm$0.030    & This paper (adopted, Sect. \ref{sec:stellarmass}) \\
Mass (M$_{\odot}$)       &    0.92$\pm$0.06    & This paper (SED, Sect. \ref{sec:sed}) \\
Radius (R$_{\odot}$)     &    0.875$\pm$0.025    & This paper (SED, Sect. \ref{sec:sed}) \\
Luminosity (L$_{\odot}$) &   0.581$\pm$0.035    & This paper (Sect. \ref{sec:stellarmass}) \\
Age  (Myr)               &    1200$\pm$200 & This paper (Sect. \ref{sec:age}) \\
$i_{\star}$ (deg)     &    $\ge$54    & This paper (Sect. \ref{sec:stellarmass}) \\
         \noalign{\smallskip}
         \hline
      \end{tabular}
\end{table}

\section{Planet validation}\label{sec:planet validation}

An alert regarding a planet candidate around TOI-4515 was released by the TESS Mission on  2021 October 21: indeed, the SPOC pipeline (\citealt{2016SPIE.9913E..3EJ}) at NASA Ames Research Center identified a candidate exoplanet with a period of 15.27 days. 
We used our independent data reduction of the FFIs to confirm the transits in the light curves and their planetary nature, following the approach described in \citet{Nardielloetal2020}. First, we modeled and removed the stellar variability by interpolating to each light curve a fifth-order spline defined over a grid of knots at intervals of 13 hours. We extracted the transit least squares (TLS) periodogram (\citealt{2019A&A...623A..39H}) of the light curve obtained by combining all the light curves after the suppression of the stellar activity. We found a peak in the TLS periodogram at $P \sim 15.27$~d with a strong signal detection efficiency (SDE$\sim 24$), confirming the presence of transit signals in the light curve. 

\begin{figure*}
  \centering
  \includegraphics[bb=14 152 575 719, width=0.95\textwidth]{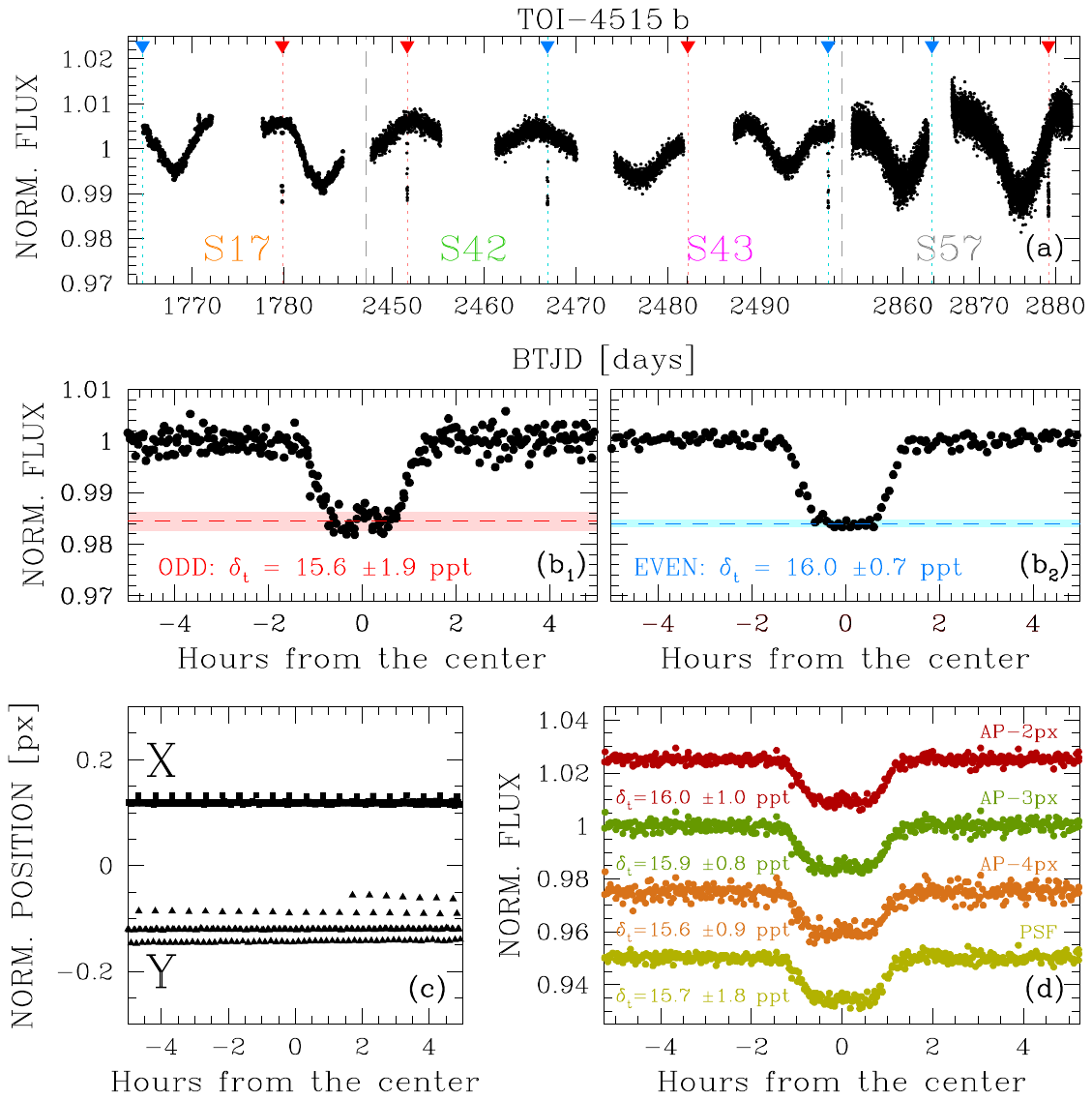}
  \caption{Overview of the vetting of TOI-4515~b. Panel (a) shows the normalized light curve obtained from the long-cadence FFIs: red and blue triangles indicate the position of odd and even transits, respectively. BTJD is the Barycentric TESS Julian date, the time stamp measured in BJD, but offset by 2457000.0, i.e., BTJD = BJD - 2457000.0.
 Panels (b) are the odd and even transits folded by using the period of 15.27 d: the transit depths are in agreement between the odd and even transits, excluding the eclipsing binary nature of the target. Panels (c) shows the X- and Y-normalized positions of the stars (stellar positions subtracted by the mean stellar position) on the image phase-folded with the period of the candidate exoplanet. Squares represent X and triangles Y. No correlation with the transits is observed. Panel (d) illustrates the phased long-cadence light curves obtained with different photometric apertures: the mean transit depth is the same within the errors, confirming that the transit signals are not due to contaminants.  \label{toi4515vet}}
\end{figure*}

Figure~\ref{toi4515vet} illustrates the vetting tests we performed to validate the planetary nature of the candidate: from the folded the odd and even transits we demonstrate that, within the errors, the transit depths are in agreement, thus we can rule out the possibility that the object is an eclipsing binary with unequal components (panel (b)). We note a bump in the odd transits, but investigating this feature, we see that this bump is only clearly visible in one of the odd transits (the last one in time, and the most densely sampled), implying the possibility of the presence of a stellar spot during the observations. 
We also exclude any correlation between the X- and Y-positions of the star in the images and the transit signals, as demonstrated in panel (c), and prove that there is no dependence between the transit depth and the photometric apertures available in our data reduction (panel (d)), minimizing the probability that the transit is due to a contaminating neighbor. The results of our data-validation tests are in agreement with those conducted by the SPOC.

\begin{figure}
  \centering
  \includegraphics[width=0.5\textwidth]{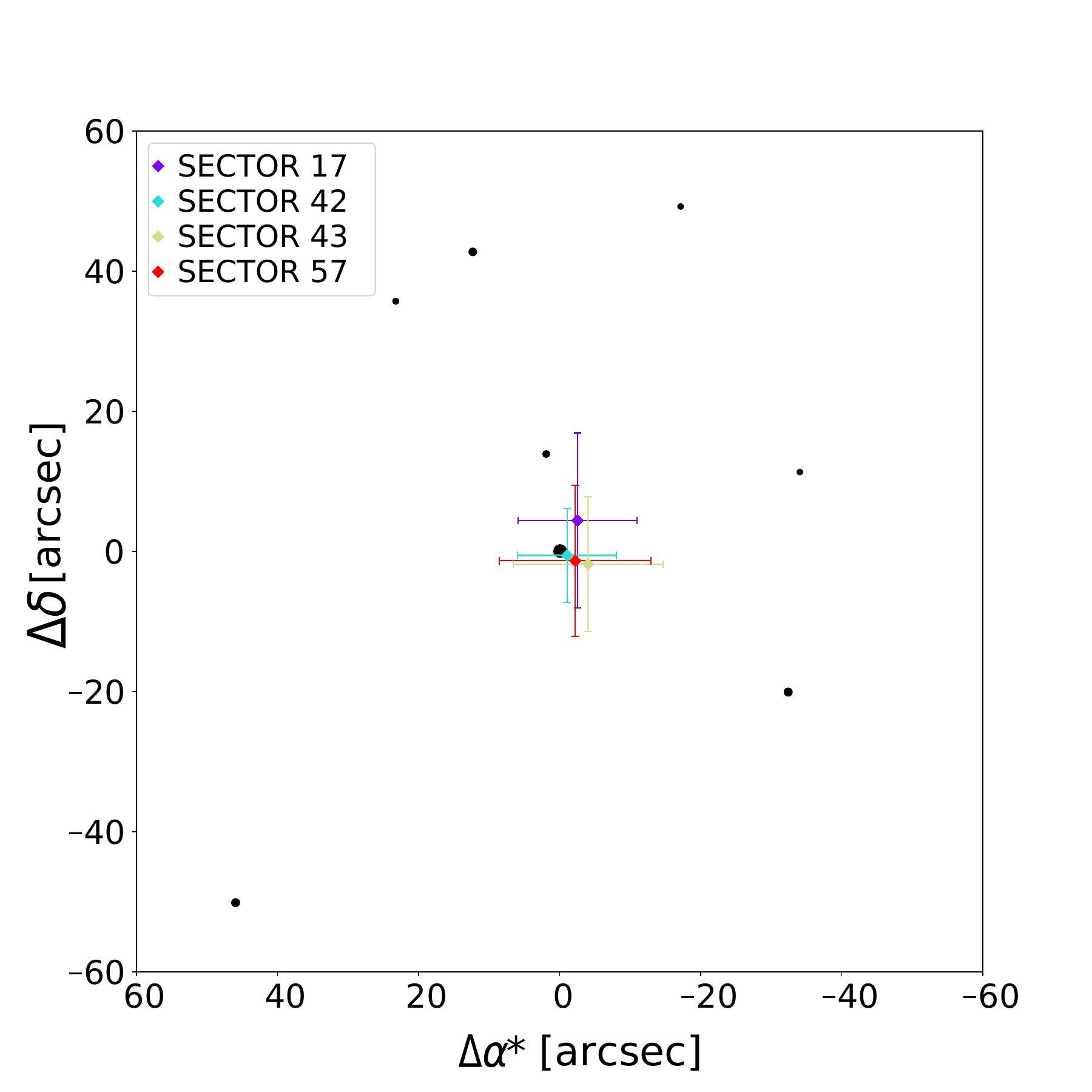}
  \caption{Analysis of the in- and out-of-transit centroid of TOI-4515~b. Within the errors, the centroid coordinates coincide with the target position (located in (0,0)). \label{centroid}}
\end{figure}

We performed the analysis of the in- and out-of-transit centroid following the procedure described by \citet{Nardielloetal2020} as a final check to verify that the transit is on TOI-4515 and is not associated with a neighboring source. The results are reported in Figure~\ref{centroid}: each mean centroid, calculated for each sector, is in agreement within the errors with the position of TOI-4515.
To ensure that our candidate is not a false positive (FP), we used the \texttt{VESPA} software \citep{Morton_2012,2015ascl.soft03010M} as a conclusive verification step. We followed the procedure outlined in \cite{2022MNRAS.516.4432M}, which takes into account the major concerns highlighted in \cite{2023RNAAS...7..107M} and allows us to get reliable outcomes while using \texttt{VESPA}. We used our detrended long-cadence light curve (see Sect. \ref{sub:tess}), which we flattened using \texttt{wotan} and then phase folded. We find a 100\% probability of having a Keplerian transiting companion orbiting TOI-4515, while the probability of an FP is very low, that is, approximately on the order of 1 x 10$^{-9}$.

\section{Planetary system analysis}
\label{sec:Planetary system analysis}
In order to retrieve the planetary system parameters, we performed a joint fit with RV and transit data using the package {\tt PyORBIT}\footnote{Available at \url{https://github.com/LucaMalavolta/PyORBIT}} (\citealt{Malavoltaetal2016,Malavoltaetal2018}). We took into account the effects of stellar activity and astrophysical contaminants. The transit modeling relies on the package \texttt{batman} \citep{Kreidberg2015} with the addition of a local polynomial trend $for each transit$. We also added a Gaussian process (GP) in the RV fit using the package \texttt{george} \citep{Ambikasaranetal2015} in order to fit the stellar activity with a quasi-periodic kernel as defined by \cite{Grunblattetal2015}, with $h$ representing the amplitude of the correlations, $\theta$ the rotation period of the star, $\omega$ the length scale of the periodic component (related to the size evolution of the active regions), and $\lambda$ the correlation decay timescale.

We modeled the TESS, KeplerCam, and LCO CTIO light curves including the following parameters: the time of first inferior conjunction $T_c$, the orbital period $P$, the eccentricity $e,$ and argument of periastron $\omega$ with the parametrization from \citealt{Eastmanetal2013} ($\sqrt{e}\cos\omega$,$\sqrt{e}\sin\omega$), the quadratic limb darkening (LD) coefficients with  \cite{Kipping2013} parametrization, the impact parameter $b$, and the scaled planetary radius $R_{P}/R_{\star}$. 
During the short duration of a transit, stellar activity due to stellar rotation can be approximated with a quadratic trend without consequences on the derived stellar parameters \citep{benatti2019, carleo2021}. This is computationally more efficient than modeling the full light curve with a Gaussian process when only a few transits are observed, as in the case of TOI-4515b. We therefore included a polynomial trend in the modelling of each of the observed transits.
Moreover, we imposed a Gaussian prior on the stellar density using the stellar mass and radius provided in Section \ref{sec:stellarmass}, and Gaussian priors for the limb darkening coefficients obtained with the code {\tt PyLDTk}\footnote{Available at \url{https://github.com/hpparvi/ldtk}}\citep{Parviainen2015, Husser2013}. 
We increased the limb darkening errors in order to avoid significant deviations between measured and predicted limb darkening coefficients, especially in TESS light curves, as explained in \citet{PatelandEspinoza2022}. We considered the impact parameter $b$ as a free parameter (e.g., \citealt{Frustaglietal2020}).  As the possible contaminating sources are negligible (Figure~\ref{fig:tpf}), the dilution factor is not included in the fit. 

Regarding the RV data, we added offset and jitter terms in order to take into account the offsets among the three different instruments and possible systematic errors and short-term stellar activity noise. When including a Gaussian process to model the activity, we also imposed a Gaussian prior on the rotational period with the value obtained from the photometric analysis (see Table \ref{t:star_param}.). 
To explore the parameter space we used the dynamic nested sampler \texttt{dynesty}\footnote{\url{https://github.com/joshspeagle/dynesty}} \citep{Speagle2020, Koposovetal2022} to sample the parameter space, with 1000 live points. 
We tested 12 models, resulting from the combination of: {\it (i}) one or two planets; {\it (ii}) with and without GP; and {\it (iii}) no trend, a linear trend, or a quadratic trend. We investigated the trend in order to rule out possible long-period additional companions, even though the RV data do not show any evident trend. 
We also computed the Bayesian evidence log$\mathcal{Z}$ from the nested sampling in order to evaluate the quality of our fits (see Bayesian values in Table \ref{tab:modelcomp}). The difference in log$\mathcal{Z}$ between the models with no trend is not significant; a small increase in the Bayesian evidence with an increase in the complexity of the model is expected even for nonfavored models (e.g., \citealt{Fariaetal2016}). 

\begin{figure}
  \centering
  \includegraphics[width=0.45\textwidth]{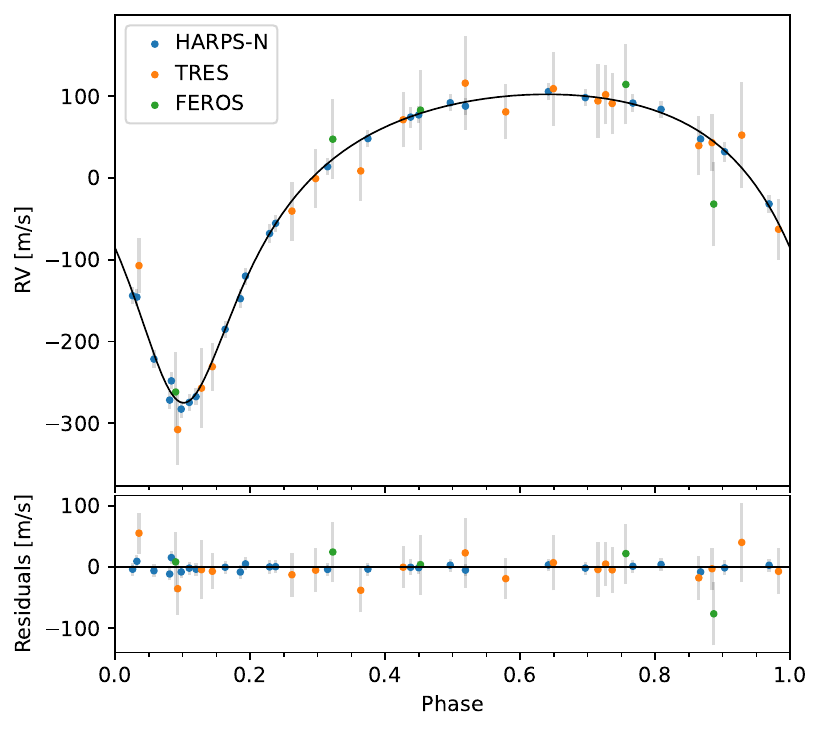}
  \includegraphics[width=0.45\textwidth]{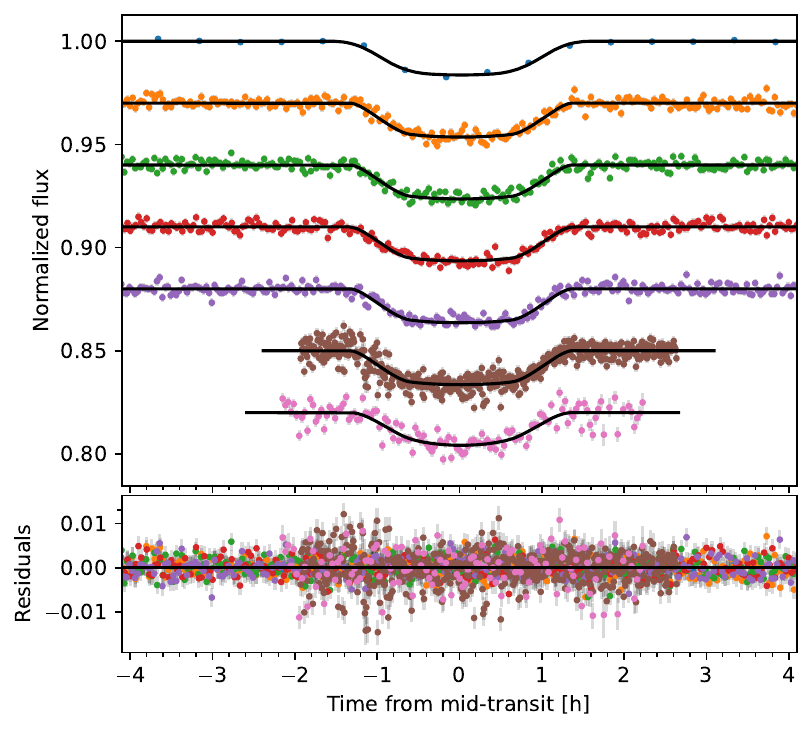}
  \caption{\textit{Upper panel}: HARPS-N (blue points), TRES (orange), and FEROS (green) RV data with the overplotted fit from the 1p model. \textit{Lower panel}: From top to bottom, TESS transits from Sectors 17 (long cadence), 42 (2 transits), 43, and 57, and LCO and KeplerCam light curves. The black line is the inferred 1p model. \label{fig:joint_fit_plot}}
\end{figure}

The planetary system parameters obtained with the two 1p and 1p+GP models with no trend are listed in Table \ref{tab:fit_params}, while the RV and transit fits with the overplotted models obtained from the model with GP are represented in Figure \ref{fig:joint_fit_plot}. We find TOI-4515\,b to have a mass of $2.03 \pm 0.05$ M$_{\rm J}$ and a radius of $1.09 \pm 0.04$ R$_{\rm J}$, with an orbital period of 15.27 d and an eccentricity of $0.46 \pm 0.01$. 
The rotational period found by the model with the GP is 15.84 d. Although this value is close to the orbital period, it is not consistent with it within 1 sigma (further discussion about the similar periods is given in Sect. \ref{sec:discussion_ecc}). However, the model without the modeling of the activity through the GP gives very similar results, meaning that the activity is not relevant in the determination of the system parameters, such as planetary mass. 
The median value of the posterior of the rotational period of the star is 15.84 days, which is apparently higher than the imposed prior, albeit well consistent within one sigma. The origin of this discrepancy may actually reside in the relatively small amplitude of stellar activity (around 10-20 m s$^{-1}$ ) compared to the planetary signal (190 m s$^{-1}$) combined with the reduced number of data collected, which causes some difficulties in modeling the activity signal in the data. In principle, the similarity between the orbital period of the planet and the stellar rotation period of the star may affect some parameter estimations, especially eccentricity and RV semi-amplitude; for example, the RV curve may be distorted by the distribution of active regions on the surface of the star, which appears in phase with the orbit. For this specific system, all of our tests confirmed that the derived planetary parameters are insensitive to the specific activity modeling. To highlight this fact, we reported in Table \ref{tab:modelcomp} the less-constrained activity model among the many tested (e.g., no training on other datasets or use of multidimensional GPs; see \citealt{nardiello2022}).

\begin{table*}
\centering
\caption{Comparison between the joint models.    \label{tab:modelcomp}}
\begin{tabular}{lccc}
\hline
Model &     No trend & Linear trend  & Quadratic trend\\
\hline
1p    & 0.00 $\pm$ 0.35 & -10.98 $\pm$  0.37 &  -324.93 $\pm$  0.35\\
1p+GP & 1.65 $\pm$ 0.35 & -10.52 $\pm$  0.38 &  -271.48 $\pm$  0.36\\
2p    & 0.92 $\pm$ 0.36 &  -9.01 $\pm$  0.38 &  -496.27 $\pm$  0.40\\
2p+GP & 1.89 $\pm$ 0.36 &  -9.55 $\pm$  0.37 &  -170.27 $\pm$  0.36 \\
\hline
\end{tabular}
~\\
~\\
  \emph{Note} --   \footnotesize The Bayesian evidence log$\mathcal{Z}$ for the 12 investigated models are listed. To improve the readability of the table, we subtracted the log$\mathcal{Z}=25071.20$ from the first model. The models with no trend are preferred.  \\
\end{table*}

\begin{table*}
  \footnotesize
  \caption{TOI-4515 parameters from the transit and RV joint fit, obtained with models 1p and 1p+GP with no trend. \label{tab:fit_params}}  
  \centering
  \begin{tabular}{lccc}
  \noalign{\smallskip}
  \hline
  \hline
  \noalign{\smallskip}
  Parameter & Prior$^{(\mathrm{a})}$ & Value$^{(\mathrm{b})}$ (1p)& Value (1p+GP)  \\
  \noalign{\smallskip}
  \hline
  \noalign{\smallskip}
  \multicolumn{3}{l}{\emph{ \bf Model Parameters }} \\
    Orbital period $P_{\mathrm{orb}}$ (days)  & $\mathcal{U}[15.25, 15.28]$   & $ 15.266447 \pm 0.000013 $ & $ 15.266446 \pm 0.000013 $\\
    Transit epoch $T_0$ (BJD - 2,450,000)  & $\mathcal{U}[9451.60, 9452.65]$  & $ 9451.62190 \pm 0.00022 $ & $9451.62191 \pm 0.00022$\\
    $\sqrt{e} \sin \omega_\star$ &  $\mathcal{U}(-1,1)$ & $ 0.119 \pm 0.023  $  &   $0.119_{-0.024} ^ {+0.025}$\\
    $\sqrt{e} \cos \omega_\star$  &  $\mathcal{U}(-1,1)$ & $-0.667_{-0.007} ^ {+0.008}$ & $-0.668_{-0.007} ^ {+0.008}$\\
    Scaled planetary radius $R_\mathrm{p}/R_{\star}$ &  $\mathcal{U}[0,0.5]$ & $0.130 \pm 0.001$   & $0.130 \pm 0.001$\\
    Impact parameter, $b$ &  $\mathcal{U}[0,1]$  & $ 0.773 _{-0.011} ^ {+0.010} $ & $0.772_{-0.011} ^ {+0.010}$\\
    Radial velocity semi-amplitude variation $K$ (m s$^{-1}$) &  $\mathcal{U}[0,300]$ & $ 193.2_{-3.6} ^ {+3.5} $  & $ 191.3 \pm 4.1 $\\ 
    \hline
    \multicolumn{3}{l}{\textbf{Derived parameters}} \\
    Planet radius ($R_{\rm J}$)  & $\cdots$ & $ 1.086  _{-0.038} ^ {+0.039} $  & $1.086  \pm 0.039$\\
    Planet radius ($R_{\oplus}$)  & $\cdots$ & $ 12.17  \pm 0.43 $  & $ 12.17  \pm 0.43 $\\
    Planet mass ($M_{\rm J}$)  & $\cdots$ & $2.026 \pm 0.047  $   & $2.005 \pm 0.052 $ \\
    Planet mass ($M_{\oplus}$)  & $\cdots$ & $644 \pm 15 $    & $637 \pm 17 $\\
    Eccentricity $e$  & $\cdots$ & $0.460 \pm 0.007$   & $0.461\pm 0.007$\\
    Scaled semi-major axis $a/R_\star$   & $\cdots$ & $ 29.68 \pm 0.27 $  &$29.67 \pm 0.27 $\\
    Semi-major axis $a$ (AU)  & $\cdots$ & $ 0.118 \pm 0.001 $  &$ 0.118 \pm 0.001 $\\
    $\omega_{\rm P} $ (deg)  &  $\cdots$ &  169.9 $\pm$ 2.0 & $169.9_{-2.2} ^ {+2.1}$\\
    Orbital inclination $i$ (deg)  & $\cdots$ & $87.958 \pm 0.041 $  & $87.954_{-0.041} ^ {+0.039} $\\
    Transit duration $T_{41}$ (days) & $\cdots$ & $ 0.135 \pm 0.002$ & $ 0.135 \pm 0.002$\\
    Transit duration $T_{32}$ (days) & $\cdots$ & $ 0.066 \pm 0.003$ & $ 0.066 \pm 0.003$\\
     \hline
    \multicolumn{3}{l}{\emph{\bf Calculated parameters}} \\
    Equilibrium Temperature T$_{eq}$ (K) &  & $705\pm10$  & $705\pm10$\\
    Planetary density $\rho_P$ (g cm$^{-3}$) &  & $1.962\pm0.216$    & $1.941\pm0.214$\\
         \hline
    \multicolumn{3}{l}{\emph{\bf Other system parameters}} \\
    Jitter term $\sigma_{\rm HARPS-N}$ (\ms) & $\mathcal{U}[0,60]$ & $13.1 _{-1.9} ^{+2.4}$  & $8.1 _{-3.3} ^{+4.0}$\\
    Jitter term $\sigma_{\rm TRES}$ (\ms) & $\mathcal{U}[0,60]$ & $24 \pm 11$ & $18 _{-11} ^{+12}$ \\
    Jitter term $\sigma_{\rm FEROS}$ (\ms) & $\mathcal{U}[0,60]$ & $47.7 _{-10.0} ^{+8.4}$  &  $35 _{-21} ^{+16}$\\
    Stellar density $\rho_\star$ ($\rho_{\odot}$) &  $\mathcal{N}[1.492, 0.045]$ & $1.507 _{-0.040} ^{+0.041}$   & $1.506 _{-0.040} ^{+0.041}$\\
    Limb darkening $q_1$  & $\mathcal{N}[0.451,0.1]$ & $0.411 \pm 0.074$  & $0.413 _{-0.072} ^ {+0.065}$\\
    Limb darkening $q_2$ & $\mathcal{N}[0.121,0.1]$ & $0.082 _{-0.084} ^ {+0.081}$  & $0.081 _{-0.082} ^ {+0.079}$\\
    Limb darkening $q_1$ (LCO) & $\mathcal{N}[0.756,0.1]$ & $0.844 _{-0.076} ^ {+0.074}$  & $0.850 _{-0.069} ^ {+0.070}$\\
    Limb darkening $q_2$ (LCO) & $\mathcal{N}[0.053,0.1]$ & $0.145 _{-0.080} ^ {+0.084}$  & $0.144 _{-0.078} ^ {+0.080}$\\
         \hline
    \multicolumn{3}{l}{\emph{\bf Stellar activity GP model Parameters}} \\
    $h_{\rm HARPS-N}$  (\ms)  &  $\mathcal{U}[0, 100]$ & &$12.7_{-5.1}^{+6.9}$ \\
    $h_{\rm TRES}$  (\ms)  &  $\mathcal{U}[0, 100]$ & &$20_{-13}^{+17}$ \\
    $h_{\rm FEROS}$  (\ms)  &  $\mathcal{U}[0, 100]$ & &$51 \pm 28$ \\
    $\lambda$  (days)  &  $\mathcal{U}[5, 2000]$ & &$704 _{-516}^{+812}$ \\
    $\omega$    &  $\mathcal{U}[0.01, 1.50]$ & &$0.16 _{-0.12}^{+0.18}$ \\
    $\theta$ (P$_{\rm rot}$)  (days)  &  $\mathcal{N}[15.5, 0.3]$ & &$15.84 _{-0.24}^{+0.16}$ \\
  \hline
   \noalign{\smallskip}
  \end{tabular}
~\\
  \emph{Note} -- $^{(\mathrm{a})}$ $\mathcal{U}[a,b]$ refers to uniform priors between $a$ and $b$, $\mathcal{N}[a,b]$ to Gaussian priors with median $a$ and standard deviation $b$.\\  
  $^{(\mathrm{b})}$ Parameter estimates and corresponding uncertainties are defined as the median and the 16th and 84th percentiles of the posterior distributions.\\
\end{table*}

\section{Constraints on additional companions both nearby and distant}
\label{sec:additionalcompanions}

A significant fraction of WJ systems have been found to host additional nearby or distant planetary companions \citep{Huang2016, Bryan2016, Wu2023}. In order to constrain the presence of additional companions, we considered the various data collected for the object (photometry, RV, and imaging, Sect. \ref{sec:observations}) as well as archival astrometric data. 

\begin{figure}
  \centering
  \includegraphics[width=0.50\textwidth]{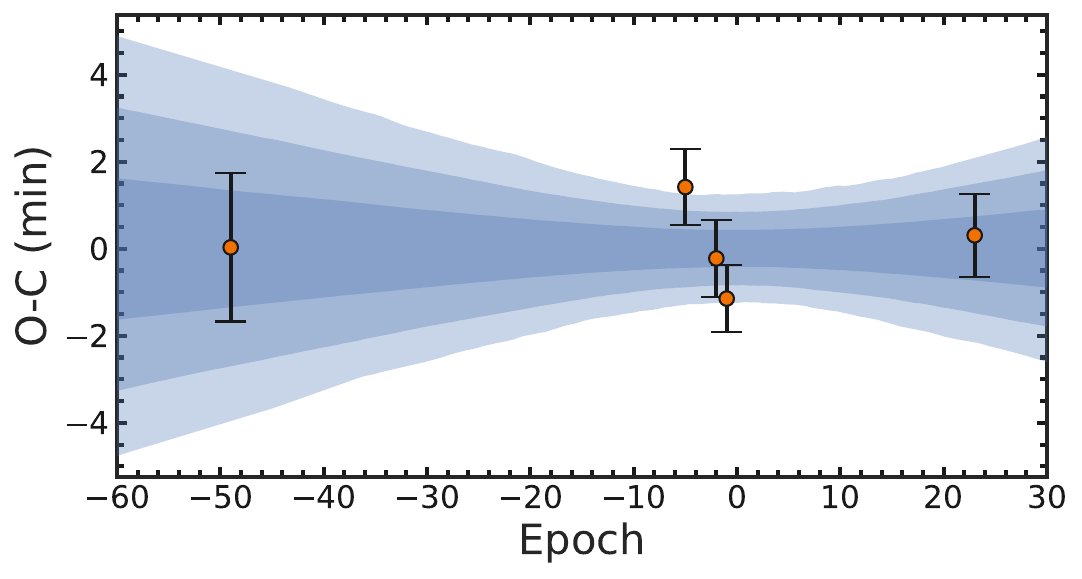}
  \caption{Observed minus calculated transit mid-times by linear ephemeris. The blue regions, from inner to outer, represent the propagation of $\pm1\sigma$, $\pm2\sigma$, and $\pm3\sigma$ errors associated with the calculated orbital period. No statistically significant TTVs are detected at levels of $\pm2\sigma$.\label{fig:oc}}
\end{figure}

To determine the precise mid-transit times and search for transit timing variations (TTVs), we conducted a global fit on the RV and transit data using \texttt{allesfitter} \citep{Max2021}. In the fitting process, we employed Gaussian priors on $T_{\rm 0}$, $P_{\rm orb}$, $\sqrt{e} \sin \omega_\star$, $\sqrt{e} \cos \omega_\star$, cos$i$, $R_\mathrm{p}/R_{\star}$, and $a/R_\star$ and transformed LD coefficients ($q_{1}$ and $q_{2}$)\footnote{The transformation equations between u-space and q-space LD coefficients are $u_{1}=2 \sqrt{q_{1}} q_{2}$, \,$ u_{2}=\sqrt{q_{1}}\left(1-2 q_{2}\right)
$.}. These parameters were sourced from Table~\ref{tab:fit_params}. For each transit, we applied an additional second-order polynomial function to account for potential trends. We employed the same method and setup as in Section~\ref{sec:Planetary system analysis} to sample the parameter space, resulting in at least 400 independent samples. Subsequently, we fit a linear ephemeris to the transit mid-times using the Markov chain Monte Carlo (MCMC) method. We optimized the reference epoch to minimize the covariance between $T_{\rm 0}$ and $P_{\rm orb}$. The resulting transit mid-times and their deviations from the linear ephemeris are listed in Table~\ref{tab:oc}. The residuals between the observed and predicted transit mid-times, as determined by the linear ephemeris, are shown in Figure~\ref{fig:oc}. There is no significant TTV signal, as all residuals with errors are consistent with zero deviation from the linear ephemeris within a $2\sigma$ confidence level.

\begin{figure}
  \centering
  \includegraphics[width=0.45\textwidth]{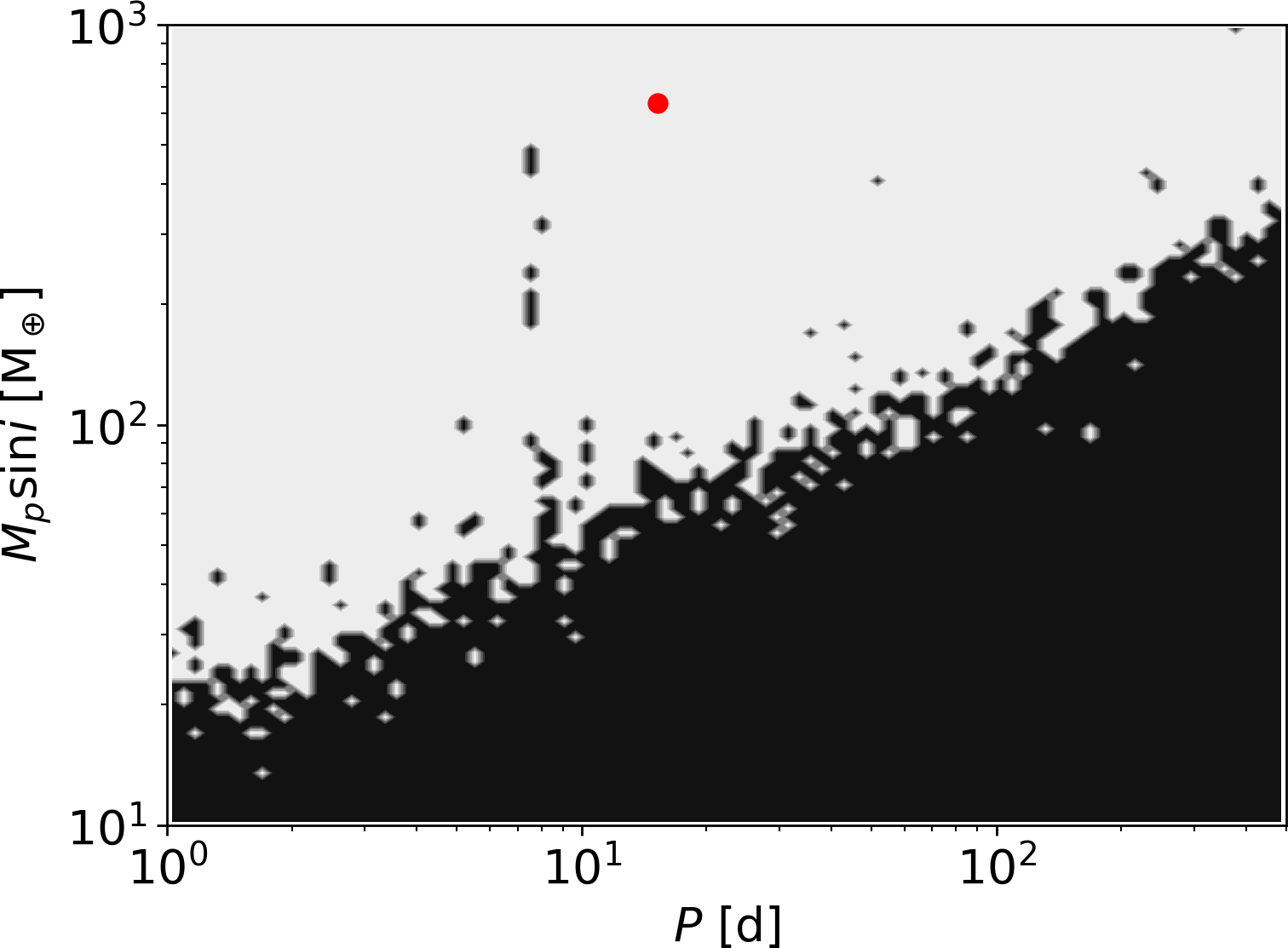}
  \caption{Detection function map of the RV time series of TOI-4515. The gray part corresponds to the area in the period--minimum mass space, where additional signals could be detected if present in the data, while the black region corresponds to the area where the detection probability is negligible. The red dot represents TOI-4515\,b, for reference. At around $P=7.5$ d, there are some low-sensitivity spots even at large masses, which are caused by the $1/2$ aliases of $P_\text{rot}$ and $P_\text{orb}$. \label{fig:detection_map}}
\end{figure}

We also tested the presence of additional nontransiting companions, deriving the detection limits from the HARPS-N RV time series. To compute the detection limits, we adopted the Bayesian technique described in \citet{pinamontietal2022}, taking into account the results of the RV and photometric data joint modeling as priors for the orbital period, $P_{\mathrm{orb}}$, and transit epoch, $T_0$, as these would be difficult to precisely constrain without the transit analysis. The resulting detection map is shown in Figure \ref{fig:detection_map}.
The detection map shows how it is currently impossible to constrain the presence of additional sub-Neptune companions ($M_p \sin i < 20$ $M_{\oplus}$), even at very short periods. We instead can exclude the presence of Saturn-mass planets at periods shorter than $40 \pm 5$ d.

\begin{table}
\caption{Optimized epochs, transit mid-times, their uncertainties, and deviations from the linear ephemeris for each transit.\label{tab:oc}}
\begin{tabular}{llll}
\hline
$N$ & $t_0$ (BJD)     & $\sigma_{t_0}$ & $\Delta_{\mathrm{linear}}$ (min) \\
\hline
-49 & 2458779.8987621 & 0.00118        & -0.01                            \\
-5  & 2459451.6237509 & 0.00060        & 1.40                             \\
-2  & 2459497.4219784 & 0.00062        & -0.24                            \\
-1  & 2459512.6877921 & 0.00054        & -1.16                            \\
23  & 2459879.0837258 & 0.00066        & 0.31             \\ \hline            
\end{tabular}
\end{table}

\begin{figure}
  \centering
  \includegraphics[width=0.50\textwidth]{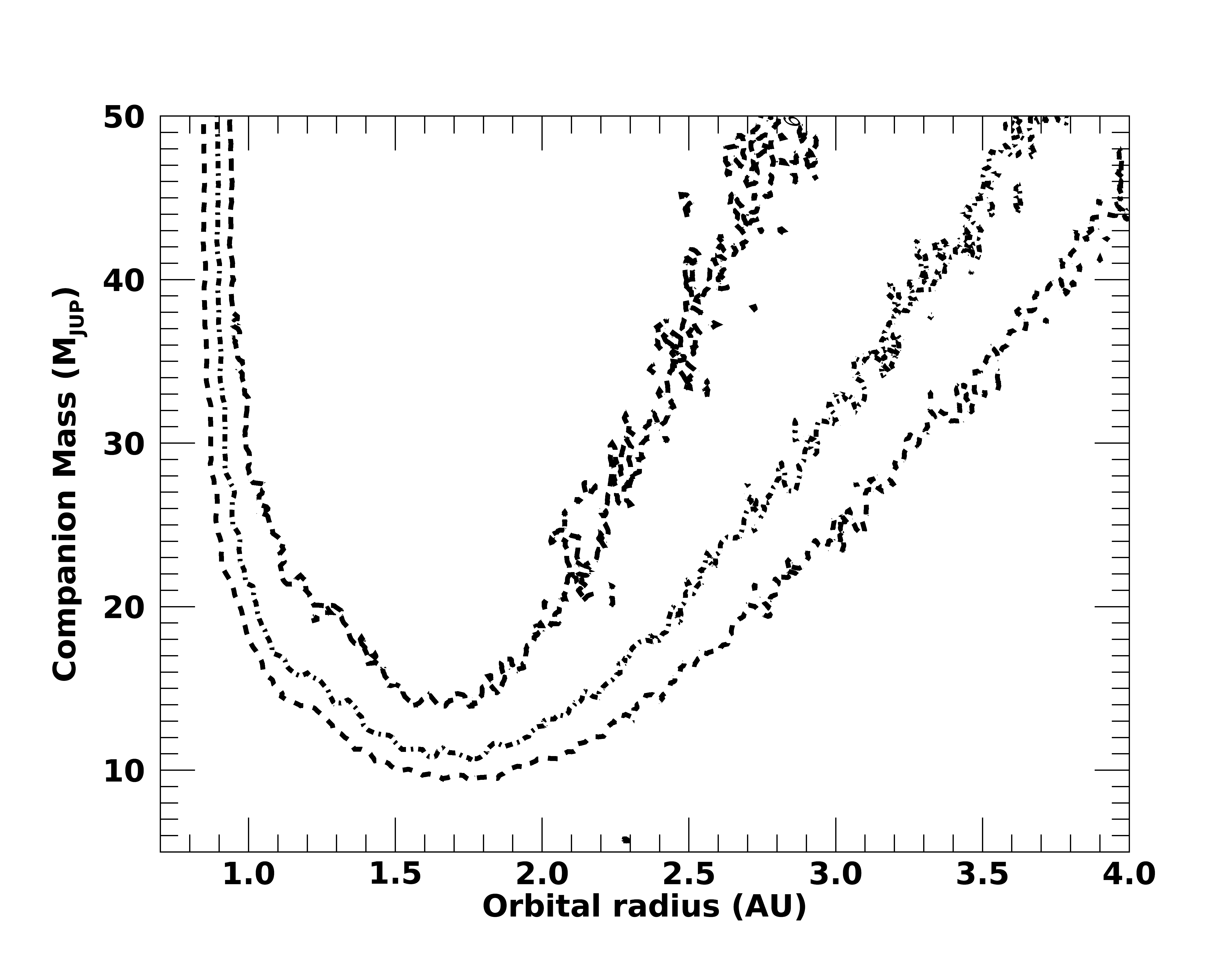}
  \caption{Limits on the presence of companions around TOI-4515 from the Gaia RUWE. The short-dashed, dashed-dotted, and long-dashed correspond to probabilities of 90\%, 95\%, and 99\%, respectively, of getting a RUWE larger than the observed value.\label{fig:ruwe}}
\end{figure}

Moving to slightly larger separations, we simulated the presence of companions able to reproduce the observed renormalized unit weight error (RUWE) from Gaia DR3 (1.09), following the procedure described in \citet{blancopozo2023}. Brown dwarf companions of various masses are ruled out in the range 1-4 au, as shown in Figure \ref{fig:ruwe}.

\begin{figure}
  \centering
  \includegraphics[width=0.50\textwidth]{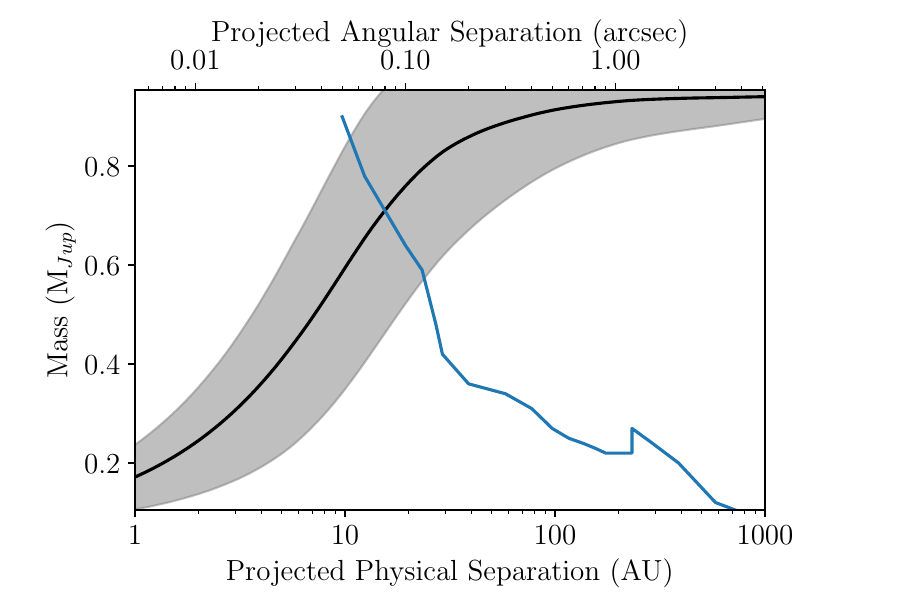}
  \caption{Constraints on the presence of companions from direct imaging and Gaia-Tycho2 proper motion differences. The detection limits from direct imaging, combining various datasets at different projected separations, are shown  as a blue continuous line (the $\textless$ 200 AU region of the blue line comes from the speckle interferometry). The continuous black line  shows the expected mass and projected separation of companions compatible with the nominal Gaia-Tycho2 proper motion difference, while the shaded gray area shows the 1 $\sigma$ limits considering realistic distributions of orbital parameters.   \label{fig:di_limits}}
\end{figure}

We also considered the imaging datasets described in Sect. \ref{sec:observations} and Gaia DR3.
No stellar companions (comoving objects) were found in Gaia DR3 within 120 arcsec (more than 20000 au at the distance of the star). The typical detection limits of close companions from Gaia are taken from \citet{brandeker2019}. The detection limits from Sect. \ref{sec:imaging} and Gaia were transformed into mass limits using the \citet{baraffe2015} models. The combined detection limits are shown in Figure \ref{fig:di_limits}. The allowed space for undetected companions is rather large, even in the stellar regime. 

Finally, we checked for indications of the presence of companions from the presence of differences in the proper motion at various epochs. As our target is not included in the Hipparcos catalog, we considered long-baseline catalogs such as Tycho2 \citep{tycho2}, PPMXL \citep{ppmxl}, and UCAC5 \citep{ucac5}, and compared the long-term proper motions from these catalogues with the short-term proper motions from Gaia DR3.
Marginally significant differences are found for Tycho2 and PPMXL catalogs.
We estimated the mass and projected separation for companions that could be responsible of the Gaia-Tycho2 proper motion difference using the COPAINS code \citep{copains}. 
Figure \ref{fig:di_limits} shows the results considering appropriate distributions of the orbital parameters.
However, we caution that, while formally significant at 2-3 $\sigma$, the observed difference may also be due to systematic uncertainties in the proper motions in pre-Gaia catalogs, as described, for example, by \citet{lindegren2016} and \citet{shi2019}. Therefore, we do not consider this as conclusive evidence of the presence of a companion with the characteristics shown in Figure \ref{fig:di_limits}. We also note that, at a separation of larger than about 30 au, the presence of a companion with the characteristics expected for the nominal Gaia-Tycho2 proper motion difference is ruled out by the imaging data.

We conclude that available data do not support the presence of additional companions, although the current detection limits allow for low-mass companions even in the stellar regime at separations of greater than about 10 au.

\section{Discussion}\label{sec:discussion}

The discovery of TOI-4515\,b, a WJ with an orbital period of $\sim15$ days, a mass of two Jupiter masses, and an eccentricity of $0.46$, orbiting a solar mass star with a metallicity of $0.05$, offers insights into the diverse formation and dynamical histories of WJs, and sheds light on the mechanisms responsible for the excitement of their eccentricities.
Considering the slightly metal-rich nature of its host star  ($\sim$ 10\% above solar metallicity), it is a possibility that multiple gas giants formed within the system \citep{Fischer2005, Wu2023}. After the gaseous disk dissipated, these gas giants may have experienced interactions, such as planet--planet scattering \citep{Anderson2020} or secular interactions \citep{Dawson_chiang2014, PetrovichTremaine2016, Dongetal2014, naoz2016}, that led to the excitation of TOI-4515\,b's eccentric orbit. Such dynamical scenarios are supported by the analysis of the normalized angular momentum deficit (NAMD) of TOI-4515\,b \citep{turrini2020,turrini2022,carleo2021}. In the case of single-planet systems, the NAMD can be expressed solely as a function of the orbital eccentricity and the spin-orbit misalignment, where neglecting the unknown contribution of the latter provides a lower bound to the dynamical excitation of the system. This lower bound to the TOI-4515\,b NAMD is $0.11$, which is about two orders of magnitude higher than the NAMD of the Solar System ($1.3\times10^{-3}$, \citealt{turrini2020}). This high value is clearly in the regime of intense chaotic evolution \citep{turrini2022} and is suggestive of catastrophic collisional events \citep{rickman2023}.

Further clues as to the dynamical past of TOI-4515\,b are supplied by its orbital and physical characteristics. First, the periastron of the system is too large to initiate tidal migration, suggesting that TOI-4515\,b did not undergo eccentric migration unless it is still dynamically coupled with companions that possess enough mass and that are sufficiently nearby to suppress the effects of general relativity, thereby sustaining the eccentricity oscillation of the WJ \citep{Wu2003}. Second, as no additional companion has been detected (as described in the following section), the eccentricity of TOI-4515\,b might be a relic of violent scattering events occurring in the distant past \citep{Anderson2020} that caused the dynamical or collisional removal of the original companion(s). Such a scenario is supported by the high density of the giant planet.

The density of TOI-4515\,b  is about 1.5 times that of Jupiter, which in turn is three times more metallic than its host star \citep{atreya2018} because of the accretion of planetary material together with gas during its formation \citep[e.g.,][]{alibert2018,oberg2019}. This comparison suggests that the  high density of TOI-4515\,b could arise from the ingestion of one or more of its original planetary companions during the dynamical instability that generated its eccentricity. If this is the case, the system may still be spin--orbit aligned \citep{Wang2021, rice2022tendency}, because scattering is less effective in exciting mutual inclination compared to the secular process, while collisions have damping effects on the dynamical excitation of planets \citep{chambers2001}. In this sense, measuring the Rossiter-McLaughlin effect for TOI-4515\,b (estimated to be $\sim$26 \ms from eq. 40 in \citealt{Winn2010}) would offer further insights into the dynamic history and formation processes of WJs like TOI-4515b.

\subsection{Photoevaporation modeling}
In order to investigate the hydrodynamic stability of TOI-4515\,b, we employed our model presented in \cite{Loccietal2019}  and updated for studying the evolution of planets spanning  from Jovian-sized to sub-Neptunian-sized bodies \citep{Benattietal2021,Maggioetal2022,Damassoetal2023,Naponielloetal2023}. In this work, we investigated the photo-evaporation of the planetary atmosphere using the energy-limited approximation \citep{Erkaevetal2007}, and taking into account the evolution of the X-ray and EUV luminosity \citep{Penzetal2008, Sanz-Forcada2011}. The time variation of the planetary radius, which evolves in response to both gravitational contraction and mass loss, is described using the theoretical model proposed by \cite{Fortneyetal2007}. By evaluating the Jeans escape parameter $ \Lambda = G m_{\rm H} M_{\rm p} / k_{\rm B} T_{\rm eq} R_{\rm p}$ (e.g., \citealt{Fossatietal2017}), we verified that the planet is stable against hydrodynamic evaporation at the present age, mostly because of its high mass.  More specifically, we obtained $\Lambda \sim 575$, which is much larger than the critical value of $\Lambda_c = 80$ for a significant atmospheric escape, and therefore the hydrodynamic mass loss rate should be negligible. We also investigated the history of the  planet in order to determine whether or not the mass and radius measured today could be the result of photo-evaporation at early ages (back in time to 10 Myr). We were not able to find any plausible planet configuration with $\Lambda < \Lambda_c$. This condition could be reached only assuming that the planet had a radius of more than a factor 7 larger than the present value ---keeping its mass  fixed---, but a young evaporating planet should also
be more massive and with a lower equilibrium temperature. We conclude that TOI-4515\,b was also stable against photoevaporation in the past, in spite of the greater high-energy irradiation.

\subsection{Eccentricity and companionship}\label{sec:discussion_ecc}
\begin{figure}
  \centering
  \includegraphics[width=0.999\linewidth,trim=0 0 0 8cm,clip]{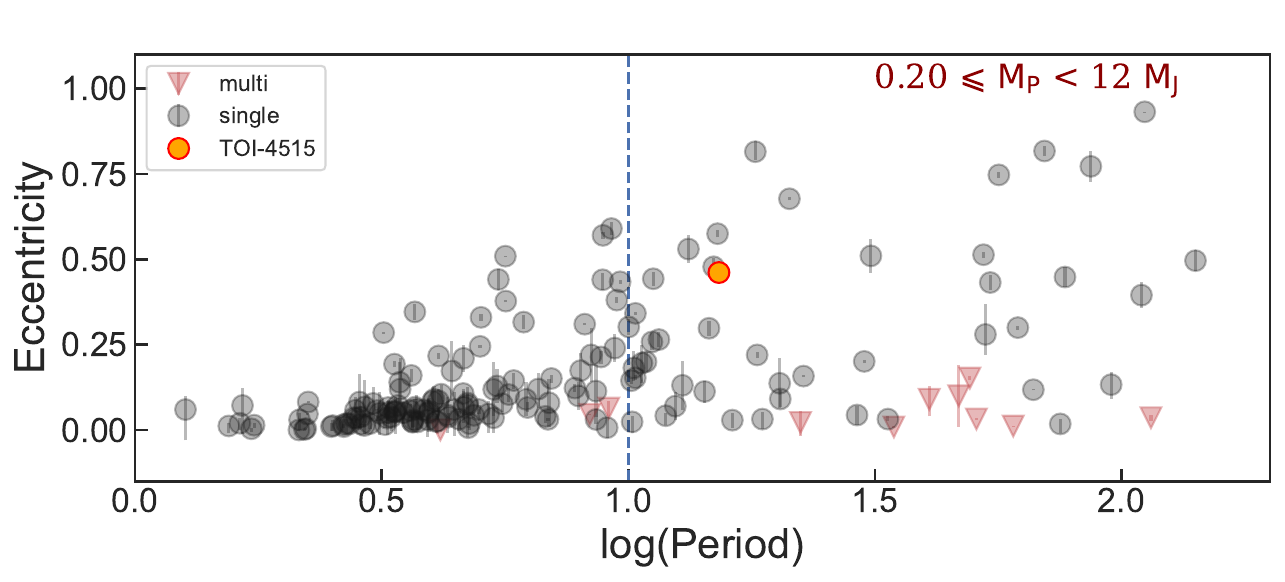}
  \caption{Distribution of eccentricities as a function of the orbital period for Jupiter-sized planets. The dashed blue line represents the ten-day boundary between HJs and WJs. The orange point represents TOI-4515\,b. The red triangles represent the planets in multi-planetary systems. Data taken as of UT 2023  September 24. \label{fig:toi4515_ecc_distrib}}
\end{figure}

To frame the orbital characteristics of TOI-4515 in the context of the population of close-in giant planets, we made use of the TEPCat catalog \citep{Southworth2011} \footnote{\url{https://www.astro.keele.ac.uk/jkt/tepcat/}} to selected planets with orbital periods within 0 < P$_{\rm orb}$ < 200 days, eccentricities with uncertainties smaller than 0.1, and planetary mass precision of better than 50\%, and mass falling in the range between 0.20 and 12 M$_{\rm J}$ (Jupiter-sized planets). The resulting planets encompass both the HJ and WJ populations and are represented in Figure \ref{fig:toi4515_ecc_distrib}, where we see that HJs with orbital periods of below 3 days are all characterized by circular orbits, which is likely due to the effective tidal dissipation that they experience throughout their lives. For increasing orbital periods, we see the appearance of giant planets on eccentric orbits alongside those on circular orbits and we see that their maximum eccentricity increases with orbital period. This trend can be intuitively explained by the rapid decrease in tidal circularization rate with increasing orbital distance (see \citealt{Jackson2008}, Eq.1), which suggests that the farther out the planet from its host star, the more likely it is that its primordial eccentricity will be preserved, at least partially. TOI-4515\,b, with its orbital period of 15 days, radius of $\sim$ 1.1 R$_{\rm J}$, mass of $\sim$ 2 M$_{\rm J}$, and eccentricity of 0.46, lies in the region of high-eccentricity WJs and its orbital eccentricity is close to the high-end tail of giant planets at similar orbital distances. The eccentricity is probably the remnant of an episode of planet--planet scattering that excited the eccentricity and also brought the planet close to the star. The tidal interaction with the star is acting to damp the primordial higher eccentricity and its circularization timescale.

In order to better understand the origin of the eccentricity of  TOI-4515\,b, we calculated the circularization timescale. An estimate of the tidal circularization timescale $\tau_{\rm e}$ (e-folding time for the decay of the eccentricity) can be made with the tidal model of \citet{Leconteetal2010}, where the modified tidal quality factors of the star $Q^{\prime}_{\rm s}$ and of the planet $Q^{\prime}_{\rm p}$ are used instead of the time lag. The approximate relationship $Q^{\prime}_{\rm s, p} \sim (3/2) k_{2 \, \rm p, s} \Delta t_{\rm s, p} n$ 
has been adopted, where $k_{2\, \rm s, p}$ is the Love number of degree 2, $\Delta t_{\rm s, p}$ is the time lag, $n=2\pi/P_{\rm orb}$ the orbital mean motion, and the subscript s or p refers to the star or the planet, respectively.
The circularization timescale depends mostly on the dissipation of the tides inside the planet with a secondary contribution from the dissipation inside the star. Adopting $Q^{\prime}_{\rm p} =10^{5}$, which is similar to the value found in the case of Jupiter by modelling the orbital evolution of the Galilean moons \citep{Ogilvie2014}, we find $\tau_{\rm e} \sim 7$~Gyr, which suggests that the eccentricity of TOI-4515b could be primordial. However, a smaller value of $Q^{\prime}_{\rm p}$ cannot be excluded given our ignorance of the internal structure of TOI-4515b and our limited understanding of tidal dissipation in giant planets. For example, with  $Q^{\prime}_{\rm p} = 10^{4}$, which was suggested in the case of Saturn by \citet{Sinclair1983}, we find $\tau_{\rm e} \sim 0.9$ Gyr, implying that the present eccentricity could require some form of excitation along the lifetime of the system.
A value of $Q^{\prime}_{\rm p}$ as small as $10^{4}$ could be observationally tested by future measurements of the night-side temperature of the planet. Specifically, as a consequence of the rather high eccentricity and low $Q^{\prime}_{\rm p}$, the power dissipated inside the planet by the tides is predicted to be of $\sim 2.6 \times 10^{20}$~W, which would imply an effective temperature of $\sim 500$~K assuming a uniform black-body irradiation from the whole surface of the body. On the other hand, the equilibrium temperature of the planet, assuming zero albedo, is $\sim 730$~K, implying that the tidal power is $\sim 24$\% of that received by the planet from its host star. Another consequence of the eccentric orbit is the pseudosynchronization of the planet, which is expected to rotate with a period of 6.2 days because of the stronger tidal interaction at periastron.
It is interesting to note that the stellar rotation period is close to the orbital period of TOI-4515\,b. This could simply be a coincidence, but it could also point to some kind of star--planet interaction as suggested by for example \citet{Lanza2022a,Lanza2022b}. The stellar synchronization timescale is longer than the age of the Universe, even assuming $Q^{\prime}_{\rm s} = 10^{5}$, which would imply an extremely strong tidal dissipation inside the star, which is not predicted by current tidal models \citep{Ogilvie2014,Barker2020}. Therefore, tides are not expected to significantly affect the stellar rotational evolution in this system. This also  applies to the obliquity of the system, which is not expected to vary over its lifetime, even assuming $Q^{\prime}_{\rm s} =10^{5}$. Therefore, a measurement of the Rossiter-McLaughlin effect can provide useful information on the formation of this system (cf. Sect. \ref{sec:discussion}).

\section{Conclusions}\label{sec:conclusion}
In this paper, we present the discovery and mass determination of an eccentric WJ transiting TOI-4515 ( TYC 1203-01161-1, TIC 456862677) and observed by \textit{TESS} in Sectors 17, 42, 43, and 57. We collected photometric (TESS, WASP, KeplerCam, CALOU, LCOGT), spectroscopic (HARPS-N, TRES, FEROS), and high-contrast imaging (SOAR, NESSI) data and constrained most of the stellar and planetary parameters. TOI-4515 is a relatively young star with an age of 1.2$\pm$0.2 Gyr and an effective temperature of about 5400 K, falling in the G-type category. We obtained a stellar rotational period of 15.5 $\pm$ 0.3 days from a combined TESS and WASP analysis. The giant planet has a mass of $2.005 \pm 0.052 $M$_{\rm J}$, a radius of $ 1.086  \pm 0.039 $ M$_{\rm J}$, and an orbital period of $15.266446 \pm 0.000013 $ days. Its eccentricity of $0.461 \pm 0.007$ places it among the sample of WJs with eccentric orbits. The combination of the eccentric orbit and a large periastron suggests planet--planet interactions, but from the available data, the presence of additional companions is not supported. However, the loss of primordial planetary companions is consistent with the high NAMD and high density of TOI-4515\,b, which jointly suggest a violent dynamical past characterized by planetary collisions. Additional data, such as imaging with the extreme-AO instruments and/or measurement of the Rossiter-McLaughlin effect, are needed in order to make major improvements and further inferences as to the architecture of the system.

\begin{acknowledgements}
This work has been supported by the PRIN-INAF 2019 "Planetary systems at young ages (PLATEA)" and ASI-INAF agreements n. 2018-16-HH.0 and 2021-5-HH.0. 
This work made use of \texttt{tpfplotter} by J. Lillo-Box (publicly available in www.github.com/jlillo/tpfplotter), which also made use of the python packages \texttt{astropy}, \texttt{lightkurve}, \texttt{matplotlib} and \texttt{numpy}.

Some of the observations in this paper made use of the NN-EXPLORE Exoplanet and Stellar Speckle Imager (NESSI). NESSI was funded by the NASA Exoplanet Exploration Program and the NASA Ames Research Center. NESSI was built at the Ames Research Center by Steve B. Howell, Nic Scott, Elliott P. Horch, and Emmett Quigley.

This work makes use of observations from the LCOGT network.

This research has made use of the Exoplanet Follow-up Observation Program (ExoFOP; DOI: 10.26134/ExoFOP5) website, which is operated by the California Institute of Technology, under contract with the National Aeronautics and Space Administration under the Exoplanet Exploration Program.

We acknowledge the use of public TESS data from pipelines at the TESS Science Office and at the TESS Science Processing Operations Center.

Resources supporting this work were provided by the NASA High-End Computing (HEC) Program through the NASA Advanced Supercomputing (NAS) Division at Ames Research Center for the production of the SPOC data products.

Funding for the TESS mission is provided by NASA's Science Mission Directorate. 

KAC acknowledges support from the TESS mission via subaward s3449 from MIT.

D.D. acknowledges support from the NASA Exoplanet Research Program grant 18-2XRP18\_2-0136.

Based in part on observations obtained at the Southern Astrophysical Research (SOAR) telescope, which is a joint project of the Minist\'{e}rio da Ci\^{e}ncia, Tecnologia e Inova\c{c}\~{o}es (MCTI/LNA) do Brasil, the US National Science Foundation’s NOIRLab, the University of North Carolina at Chapel Hill (UNC), and Michigan State University (MSU).

A.J.\ acknowledges support from ANID -- Millennium  Science  Initiative -- ICN12\_009 and from FONDECYT project 1210718.

R.B.\ acknowledges support ANID -- Millennium  Science  Initiative -- ICN12\_009 and from FONDECYT project 11200751.

S.Q. acknowledges support from the TESS GI Program under award 80NSSC21K1056 and from the TESS mission via subaward s3449 from MIT.

D. D. acknowledges support from the NASA Exoplanet Research Program grant 18-2XRP18\_2-0136, and from the TESS Guest Investigator Program grants 80NSSC22K1353 and 80NSSC22K0185.

S.W. gratefully acknowledges the generous support from the Heising-Simons Foundation, including support from Grant 2023-4050.

For the purpose of open access, the authors have applied a Creative Commons Attribution (CC BY) licence to any Author Accepted Manuscript version arising from this submission.

\end{acknowledgements}

%
%

\begin{appendix}
\section{RVs}

\begin{table*}[!ht]
\caption{\label{tab:rvdata} Time series of TOI-4515 from HARPS-N, TRES, and FEROS data: Julian dates, RVs, and their related uncertainties. For HARPS-N data the \logrhk values are listed as well.}
\begin{tabular}{l|cccrc}
\hline
\noalign{\smallskip}
& \multirow{2}{*}{JD - 2450000}  &      RV  & $\sigma_{\rm RV}$  &   $\rm log\,R^{\prime}_\mathrm{HK}$ & $\sigma_{\rm log\,R^{\prime}_\mathrm{HK}}$  \\ 
     &        & (\ms) &    (\ms)   & & \\
\hline             
\noalign{\smallskip}
HARPS-N & 9561.430142  &   12994.2       &     2.9     &  -4.654      &     0.015  \\
 &   9565.348535   &     13186.8    &        2.2    &  -4.673     &      0.010  \\
 &   9566.405805   &     13194.3    &        2.6    &  -4.707     &      0.014  \\
 &   9575.424766   &     12829.9    &        3.4    &  -4.653    &       0.018  \\
 &   9579.465702  &      13152.8    &        3.5    &  -4.689   &        0.021  \\
 &   9580.432180  &      13181.7    &        7.9    &  -4.664    &       0.055  \\
 &   9581.327863  &      13201.7    &        3.3    &  -4.674    &       0.018  \\
 &   9584.383000  &      13210.6    &        2.8    &  -4.672    &       0.014  \\
 &   9585.458583  &      13209.6    &        3.1    &  -4.684    &       0.018  \\
 &   9601.362277   &     13201.2    &        2.5    &  -4.668    &       0.012  \\
 &   9624.353376  &      13119.9    &        2.4    &  -4.696    &       0.013  \\
 &   9629.347520  &      13212.9    &        2.8    &  -4.694    &       0.016  \\
 &   9633.332790  &      13148.1   &         7.3    &  -4.620    &       0.045  \\
 &   9634.337974 &       13086.4    &        2.9    &  -4.712     &      0.018  \\
 &   9772.704757  &      12943.9   &         2.2    &  -4.654    &       0.009  \\
 &   9773.707306  &      12808.8    &        2.1    &  -4.650     &      0.009  \\
 &   9774.700799  &      12917.1   &         2.6    &  -4.666     &      0.012  \\
 &   9775.705586  &      13040.4   &         6.3    &  -4.622     &      0.034  \\
 &   9788.712922  &      12818.2  &          2.9    &  -4.689     &      0.015  \\
 &   9800.721264  &      13149.7   &         2.8    &  -4.712     &      0.014  \\
 &   9803.623676  &      12882.8   &         3.3    &  -4.674     &      0.016  \\
 &   9804.575617  &      12824.3   &        3.4    &  -4.683      &     0.018  \\
 &   9805.576617  &      12941.3    &        5.2    &  -4.644     &      0.029  \\
 &   9821.641945   &     13033.7   &         3.6    &  -4.680     &      0.019  \\
 &   9833.673975   &     12965.6    &        2.1    &  -4.697     &      0.009  \\
 &   9834.550525  &      12870.1    &        3.1     &  -4.665    &       0.015  \\
\noalign{\smallskip}
\hline
\noalign{\smallskip}
TRES & 9516.681944       & -87.1 &      29.4   &   &    \\
 & 9523.776696 &        67.5 &  29.7   &   &    \\
 & 9549.730853 &        20.3 &  26.7   &   &    \\
 & 9556.715849 &        14.3 &  27.6   &   &    \\
 & 9560.680939 &        -271.7 &        20.5   &   &    \\
 & 9569.724655 &        51.6 &  31.4   &   &    \\
 & 9571.682804 &        0.00 &  29.7   &   &    \\
 & 9572.654784 &        22.2 &  61.6   &   &    \\
 & 9575.702477 &        -291.2 &        43.7   &   &    \\
 & 9581.666751 &        65.1 &  53.3   &   &    \\
 & 9582.583227 &        29.8 &  26.4   &   &    \\
 & 9583.660704 &        67.5 &  40.2   &   &    \\
 & 9584.668391 &        57.0 &  39.9   &   &    \\
 & 9771.950045 &        -104.5 &        31.1   &   &    \\
 & 9833.817876 &        -102.3 &        26.0   &   &    \\
 & 9837.813332 &        -62.7 &         28.9   &   &    \\
 & 9838.833613 &        -63.5 &         29.7   &   &    \\
 & 9849.960417 &        -303.5 &        37.4   &   &    \\
\noalign{\smallskip}
\hline
\noalign{\smallskip}
FEROS  & 9544.58457   &         12806.3   &     9.3   &  -4.637 &   0.037     \\
  & 9844.82956   &      13183.1   &     10.8    &  -4.826       &   0.099     \\
  & 9846.81782   &      13010.4   &     19.6    &  -4.227       &   0.107     \\
  & 9868.73429   &      13121.5   &     11.3   &  -4.541        &   0.044     \\
  & 9870.71811   &      13151.1   &     10.5    &  -4.745       &   0.057    \\
\noalign{\smallskip}
\hline
\noalign{\smallskip}
\end{tabular}
\end{table*}


\clearpage

\section{Corner plot}

\begin{figure*}[!ht]
  \centering
  \includegraphics[width=0.95\textwidth]{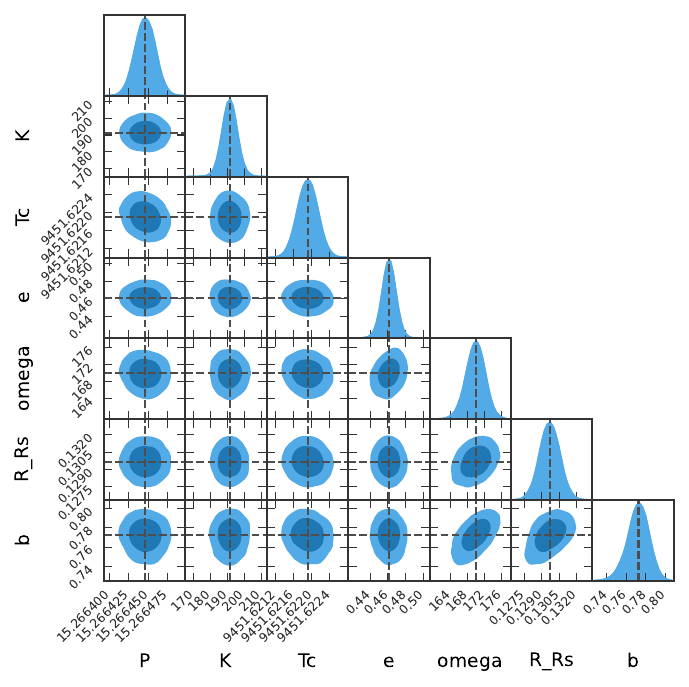}
  \caption{Corner plot of the posterior distributions for the planetary parameters obtained with the  1p+GP model. \label{fig:corner_plots}}
\end{figure*}

\end{appendix}

\end{document}